\documentclass[aps,pre,reprint,groupedaddress,notitlepage]{revtex4-1}

\usepackage{graphicx}
\usepackage{amsmath}
\usepackage{amssymb}
\usepackage{subfigure}
\usepackage[labelfont=bf,labelsep=period,justification=raggedright]{caption}

\usepackage{bibunits}
\defaultbibliography{pnmfinal}

\newcommand{\flensfree}{$G_F$\ }
\newcommand{\flenspar}{$G_{FP}$\ }
\newcommand{\flenscon}{$G_{FPC}$\ }

\begin{document}

\title{Untangling the roles of parasites in food webs with generative network models}

\author{Abigail Z.\ Jacobs$^1$}
\email[]{abigail.jacobs@colorado.edu}

\author{Jennifer A.\ Dunne$^2$}
\author{Cris Moore$^2$}
\author{Aaron Clauset$^{1,2,3}$}

\affiliation{${^1}$Department of Computer Science, University of Colorado, Boulder, CO 80309\\
${^2}$Santa Fe Institute, 1399 Hyde Park Rd., Santa Fe, NM 87501\\
${^3}$BioFrontiers Institute, University of Colorado, Boulder, CO 80303}

\begin{bibunit}

\begin{abstract}
Food webs represent the set of consumer-resource interactions among a set of species that co-occur in a habitat, but most food web studies have omitted parasites and their interactions. Recent studies have provided conflicting evidence on whether including parasites changes food web structure, with some suggesting that parasitic interactions are structurally distinct from those among free-living species while others claim the opposite. Here, we describe a principled method for understanding food web structure that combines an efficient optimization algorithm from statistical physics called parallel tempering with a probabilistic generalization of the empirically well-supported food web niche model. This generative model approach allows us to rigorously estimate the degree to which interactions that involve parasites are statistically distinguishable from interactions among free-living species, whether parasite niches behave similarly to free-living niches, and the degree to which existing hypotheses about food web structure are naturally recovered. We apply this method to the well-studied Flensburg Fjord food web and show that while predation on parasites, concomitant predation of parasites, and parasitic intraguild trophic interactions are largely indistinguishable from free-living predation interactions, parasite-host interactions are different. These results provide a powerful new tool for evaluating the impact of classes of species and interactions on food web structure to shed new light on the roles of parasites in food webs.\end{abstract}

\maketitle

\section{Introduction}
 
\begin{figure}
\includegraphics[width=3.35in]{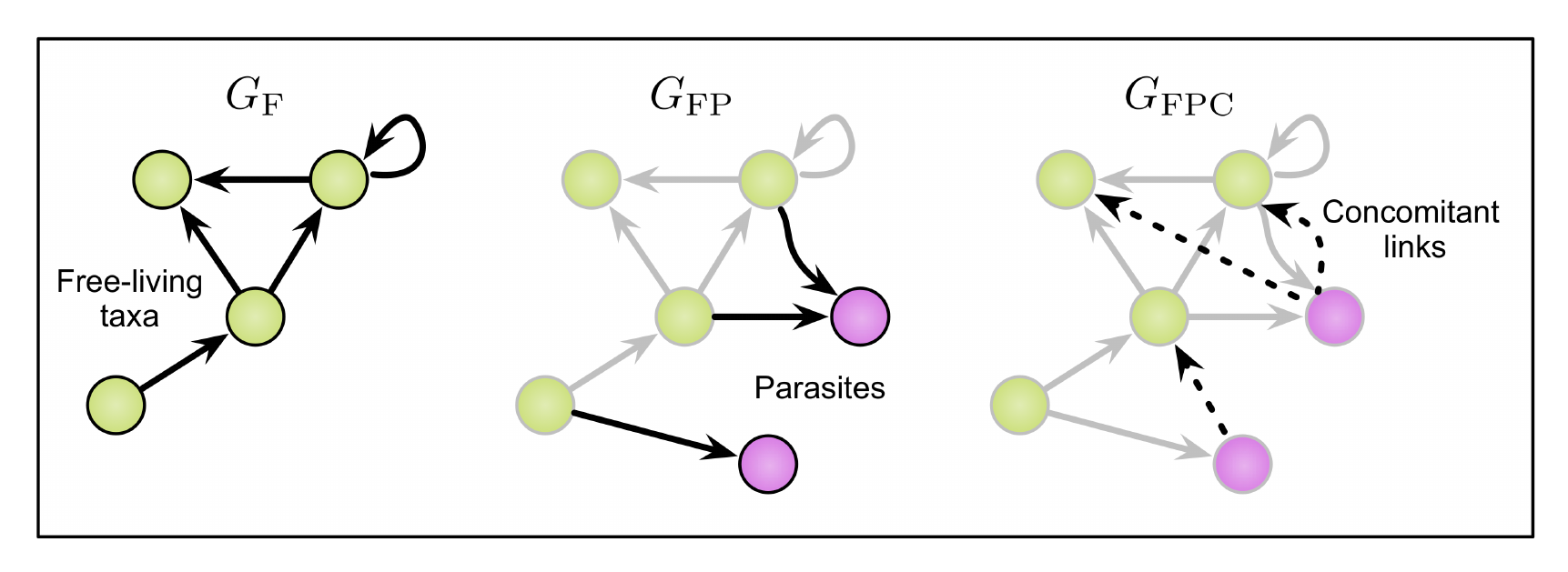}%
\caption{{\bf Three versions of a food web, showing the progressive addition of information.} \flensfree indicates the interactions among free-living species only, \flenspar adds parasites and their interactions with free-living species, and \flenscon adds concomitant links from free-living species to parasites.
\label{fig1}}
\end{figure}

\begin{figure*}
\includegraphics[width=4in]{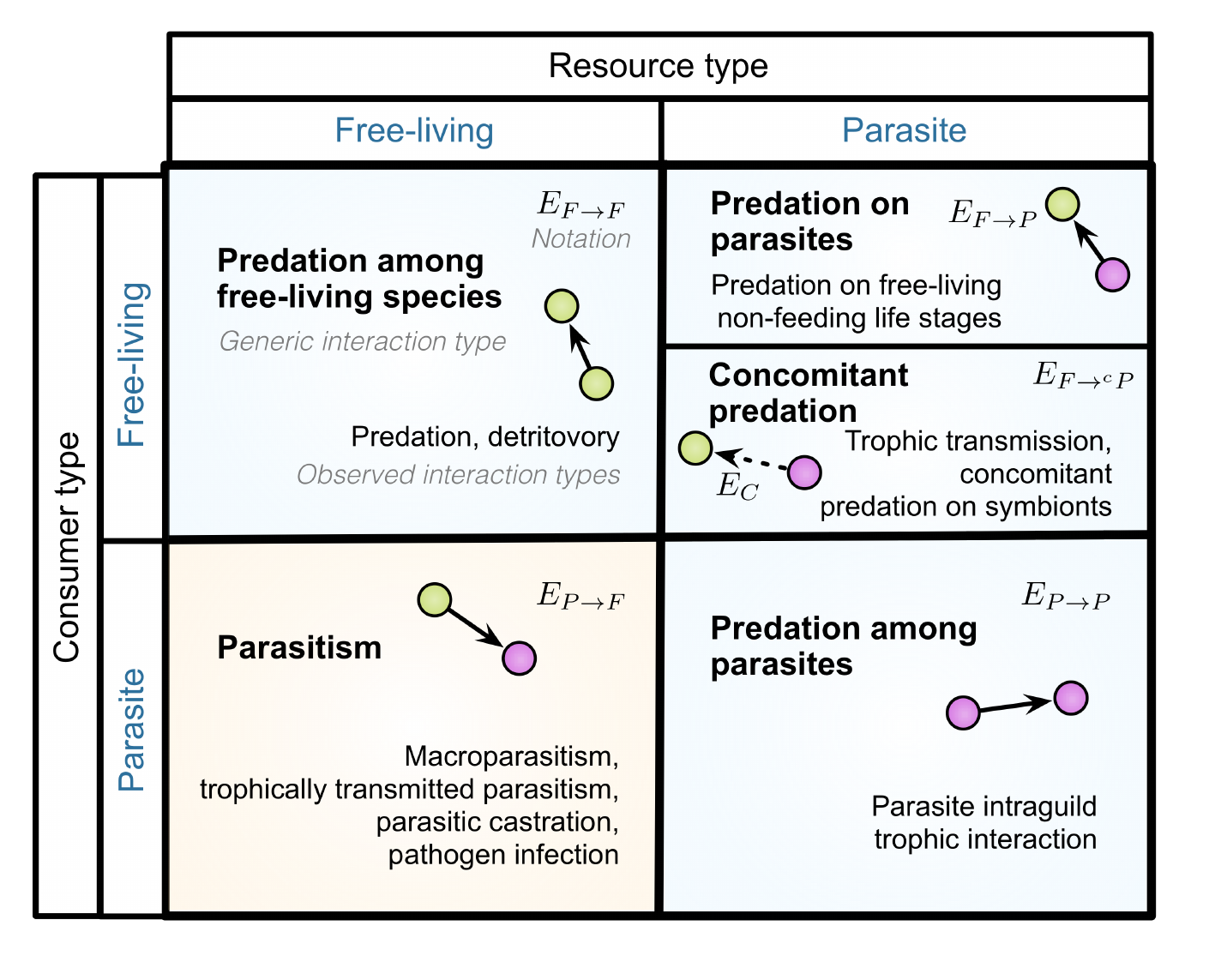}%
\caption{{\bf Interaction types.} The types discussed are paired with the precise interaction types in the data. Broadly, the types fall into two categories: predation (top row and bottom right) and parasitism (bottom left). The free-living to parasite edgeset $E_{F\rightarrow^c P}$ includes both predation on parasites $E_{F\rightarrow P}$ and concomitant predation links $E_C$, thus is in $G_{FPC}$ only.
\label{fig2}}
\end{figure*}

Ecological networks, and food webs in particular, are a useful quantitative tool for evaluating and understanding the structure and function of complex ecosystems. Most food web studies have focused on the interactions among free-living species, and have omitted diverse and ecologically important groups like parasites, which contribute to ecosystem function~\cite{hudson2006healthy}. Food web research is beginning to explicitly include parasites, but it remains unclear whether parasites and free-living species (Figure~\ref{fig1}) play distinct roles in structuring food webs, and thus whether food web theory needs to be altered to account for different types of feeding interactions (Figure~\ref{fig2}). Resolving this question would shed new light on the fundamental principles of trophic organization in ecosystems.


Several recent studies have considered this question, but substantial ambiguity remains, in part because the competing hypotheses have primarily been tested indirectly, focusing on the impact that including parasites has on standard statistical measures of food web structure. For instance, many studies have argued that parasites alter food web structure in fundamental ways~\cite{huxham1996triangles,Marcogliese-parasiteplea,para-missing-links,johnson-parasites-prey,fontaine2011ecological,kefi2012merging,britton-2013}, partly as a result of parasites’ unique characteristics like small body sizes compared to their hosts, trophic intimacy with their hosts, and often complex life cycles~\cite{huxham1996triangles,thompson2005importance,hernandez2008parasites,kuangzhang2011,fw-structurefn-biodiversity}. One proposal to explain such differences posits a distinct “inverse niche space” that parasites occupy~\cite{inverse-niche-model}, 
which allows parasites and free-living species to follow different rules of interaction.

On the other hand, one recent study~\cite{dunne-2013-parafw} showed that most of the changes to common network measures for food webs that result from adding parasites and their links are largely what we should expect simply from changing the food web’s scale by increasing the number of species $S$ and links $L$~\cite{martinez1998time}. This study noted two exceptions. First, concomitant links, the feeding links connecting predators to the parasites of their prey~\cite{johnson-parasites-prey}, appear to alter the observed frequencies of certain motifs representing interactions among triplets of species~\cite{stouffer2007evidence}.  Second, generalist parasites, which have multiple hosts, appear to have more complex trophic niches than generalist free-living predators~\cite{PNM2011}. The disagreement and accordance between this and past studies illustrates the complexity of the question of whether and how parasites alter food web structure, and demonstrate the need for statistically rigorous tools for addressing it~\cite{dunne-2013-parafw}. 

A subsequent study further investigated the distributions of motifs among free-living species, parasites, and different types of interaction links~\cite{cirtwill2015concomitant}. This study showed that parasites have unique structural roles compared to free-living species: when concomitant links were excluded, parasites have diverse roles similar to free-living consumers that have both predators and prey (i.e., intermediate taxa), but when concomitant links were included, their roles were more constrained and different~\cite{cirtwill2015concomitant}. This study also found that concomitant links represent the most structurally diverse type of interaction.

A common feature of earlier work on these questions is the indirect nature in which they tested the “null” hypothesis that parasites and free-living species follow similar rules in how they fit into food webs~\cite{williams-martinez-niche-2000}. This leaves open the possibility that parasites alter food web structure in subtle but important ways not apparent through existing approaches. In particular, most previous work has lacked either rigorous hypothesis testing or an explicit comparison of parasite interaction types to those among free-living species, or it has focused on changes in network-level statistics without controlling for confounding effects. A more direct approach would use an explicit but realistic null model of food web structure. We construct such a probabilistic null model based on the hypothesis that there is no difference in the structural roles of parasites and free-living species, but which nevertheless represents realistic food web structure. This approach can demonstrate whether a single set of rules is sufficient to simultaneously explain the interaction patterns of both parasites and free-living species. Here, we introduce a novel computational method for making just such a direct test, which we demonstrate by untangling the role of parasites within the well-studied Flensburg Fjord food web~\cite{dunne-2013-parafw,flensburg-fjord-data}.

Our approach is based on the probabilistic generalization of the empirically well-supported niche model~\cite{williams-martinez-niche-2000}, called the probabilistic niche model or PNM~\cite{PNM2011,PNM2010}. The PNM enables the inference of an underlying niche structure that explains the observed links in a food web. Specifically, the model assumes a single underlying niche space, in which each species $i$ is located at some niche location $n_i$ and probabilistically feeds on species located near location $c_i$, the center of species $i$‘s feeding range of width $r_i$. Through certain patterns on these parameters, the PNM can capture a variety of empirically supported structural features, such as hierarchical feeding, compartmental structure, and body-size determined feeding niches~\cite{cattin2004phylogenetic,pimm1980compartments,woodward2005body}. The PNM can also capture cascade structure and inverse niche structure~\cite{inverse-niche-model,cohen-newman-cascade} (see also Supporting Information S4, Table S4). Generalizing over these hypotheses, we can assess the statistical quality of the overall model as multiple types of taxa and interactions are introduced. These characteristics make the PNM an attractive and powerful method for testing hypotheses about the structural role of parasite species and their interactions in food webs.

However, fitting the PNM to an existing food web does not itself test the hypothesis that parasites and free-living species follow distinct rules for feeding. Instead, we first partition the types of interactions, e.g., predation among free-living species versus free-living predation on parasites versus concomitant links. We then test whether or not parasites and free-living species or their interactions are different by comparing models across the sequence of food webs created by adding these interaction types one at a time. If parasites follow distinct patterns, then adding parasites and their links to a free-living food web forces the model to fit a broader variety of interaction patterns, which will result in a decrease in the PNM’s goodness-of-fit. On the other hand, if parasites play similar structural roles to free-living species, adding these taxa and their links will not impact the model’s goodness-of-fit. Similarly, if concomitant links follow a distinct pattern to links among and between parasites and free-living species, adding them will result in another decrease in the PNM’s goodness-of-fit. In this way, the PNM provides a mechanism-agnostic method for detecting heterogeneities in linkage patterns as more types of interactions are added, without having to identify the particular ecological mechanism at play. 

Across different subsets of the data, we use three distinct goodness-of-fit measures to test the null hypothesis: one based on in-sample learning, one based on out-of-sample learning, and one based on statistical network feature similarity. A model that performs better under each of these measures more effectively captures the observed pattern of interaction in the food web. Conversely, if interaction types fail to be well represented, the goodness-of-fit measures will reveal this. By evaluating the null hypothesis several different ways, we increase the reliability of the overall test for differences between types of species and interaction types. 
Each of these tests requires estimating the best underlying niche structure for a food web, which is a difficult optimization problem related to finding the global maximum of a likelihood function characterized by many local optima. Previous work with the PNM~\cite{PNM2010} has used simulated annealing, which is inefficient and sensitive to initialization heuristics. Furthermore, initialization heuristics can induce a strong bias when applied to food webs containing parasites, making them unsuitable for testing the hypotheses of interest here. To resolve these technical difficulties, we use a sophisticated and more efficient optimization technique from statistical physics called parallel tempering to fit the PNM directly to the links contained in a food web, without any initialization heuristics. Compared to a number of alternative optimization techniques, this method achieves substantially better results on food web data.

We apply our method to the Flensburg Fjord food web, a single, well-resolved coastal food web, to demonstrate that this technique can address specific open questions about parasites in food webs. We use this data set, with and without parasites and with and without concomitant links, and trophically aggregated by species and over life stages~\cite{dunne-2013-parafw,flensburg-fjord-data}. Here, we focus on demonstrating how to untangle the role of parasites or other types of taxa in food webs using generative models. We intentionally leave a comparative investigation across food webs for future work, and instead emphasize the development and demonstration of these methods, which can be applied to a wide range of questions about ecological network structure. We use these methods to specifically address the question of whether parasites and free-living species exhibit statistically distinguishable interactions or niches, and whether these methods can be used to recover previously hypothesized models.

\section{Results}

\begin{figure}
\includegraphics[width=3.35in]{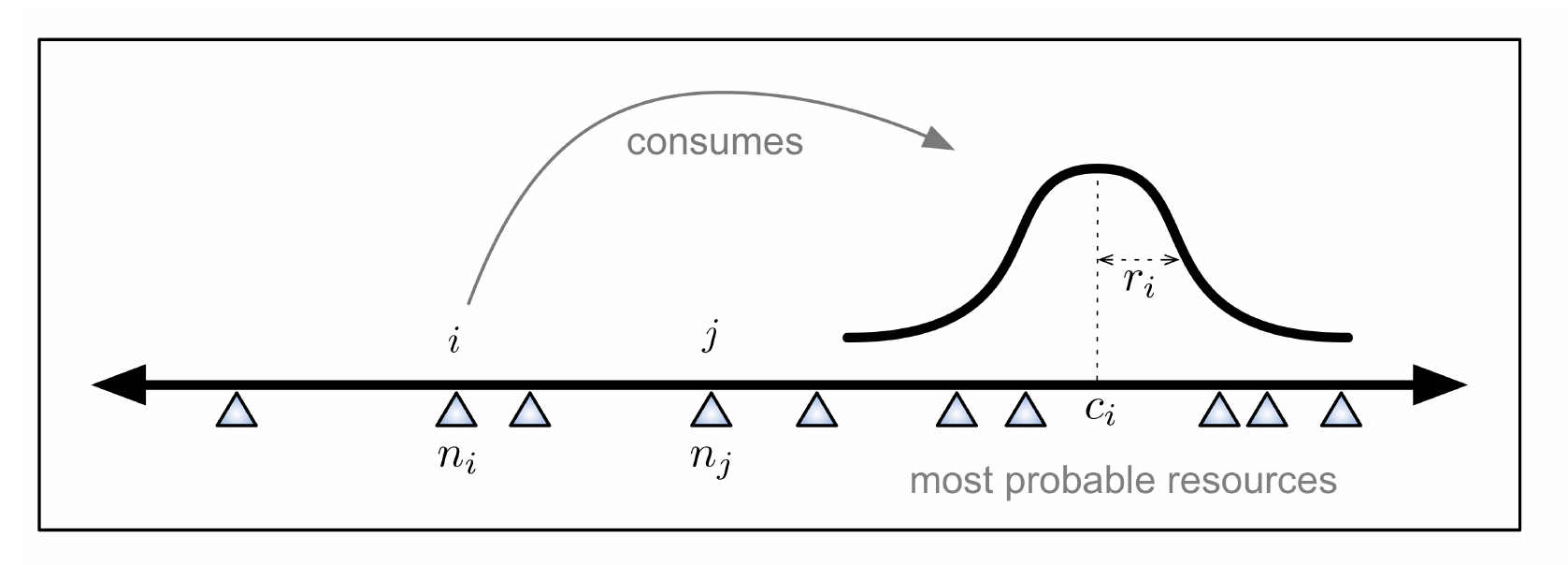}%
\caption{{\bf Schematic of the probabilistic niche model (PNM).}
Each species $i$ has some position $n_i$ in a latent niche space, represented here by a one-dimensional axis. Species $i$ consumes each other species $j$ (located at $n_j$) with a probability given by a Gaussian function centered at a preferred feeding location $c_i$ and width $r_i$. Taxa whose niche positions are nearer the preferred feeding location are consumed with higher probability.
\label{fig3}}
\end{figure}

The probabilistic niche model (PNM) is a generative model for food web structure. Each species $i$ in the food web is represented by a triplet of parameters: $n_i$, $c_i$ and $r_i$. The settings of these parameters for each species represent the underlying niche structure of the model. A directed consumer-resource link from species $i$ to species $j$ is generated with probability given by a generalized Gaussian function (Figure~\ref{fig3}). Thus, the PNM defines a parametric probability distribution over all possible food web structures, and the particular settings of the parameters determine with types of structures are more or less likely to be generated. In observed food web data, the underlying niche structure is unknown, while the links are observed. For a given set of species and feeding links, we estimate the niche structure---i.e., the three parameters for each species---via maximum likelihood. 

\begin{figure}
\includegraphics[width=3.35in]{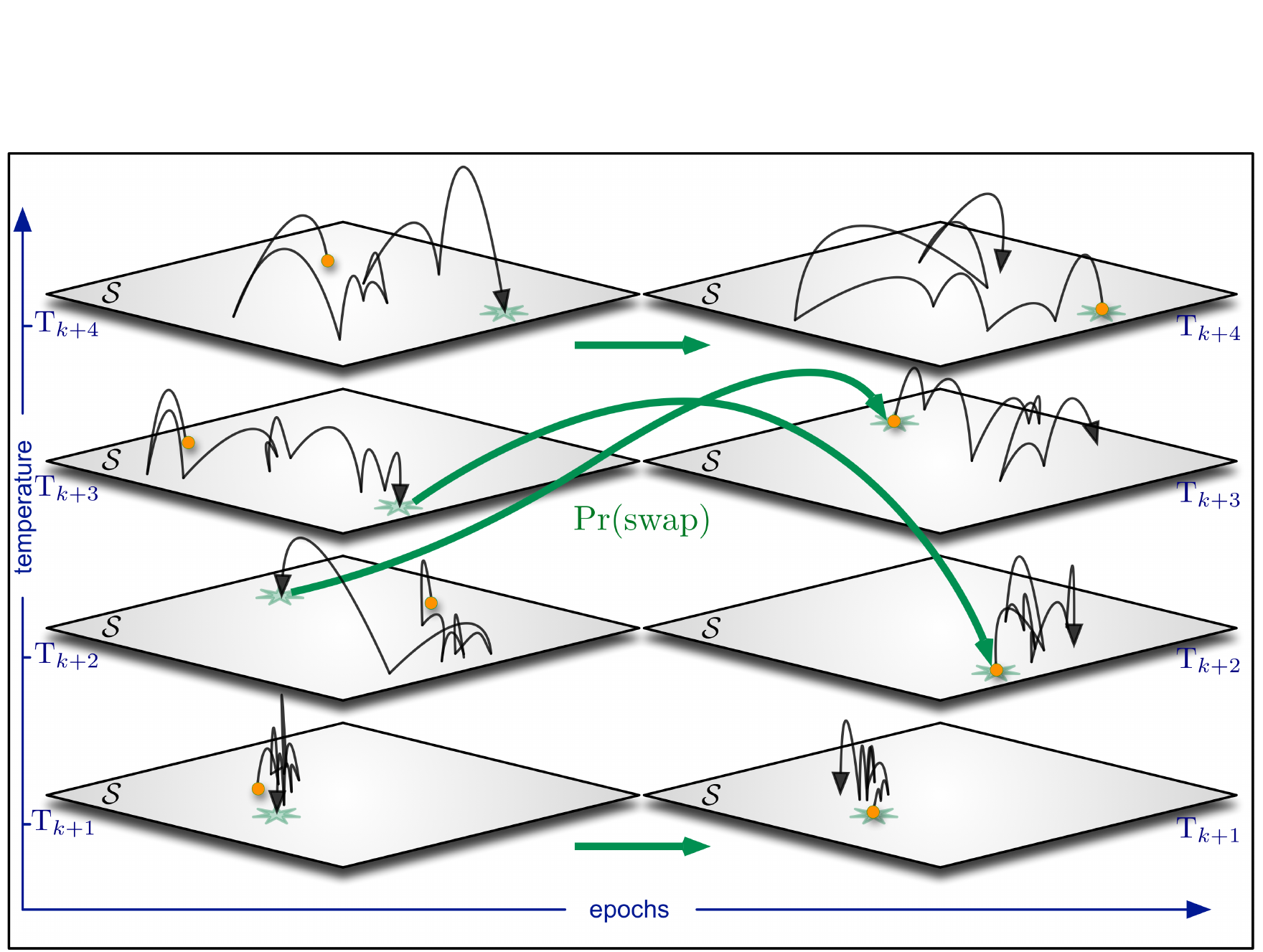}
\caption{{\bf Schematic of the parallel tempering method for fitting the model.} 
In parallel tempering, individual Markov chain Monte Carlo simulations run in parallel but at different temperatures that range between uniform exploration (top, high temperature) and greedy exploration (bottom, low temperature). Chains run as usual MCMC within an epoch of time. At the end of each epoch, a uniformly random neighboring pair of chains will exchange states according to a standard Metropolis-Hastings rule. The result is an efficient combination of liberal and greedy exploration strategies for find high-scoring maxima in the search space.
\label{fig4}}
\end{figure}

The likelihood function for the PNM is known to be rugged, exhibiting many local optima. This property makes it difficult to find the maximum likelihood niche structure via standard techniques, e.g., greedy optimization or gradient descent~\cite{hastie2009elements}. To circumvent this difficulty, we used a state-of-the-art optimization technique from statistical physics called parallel tempering~\cite{paralleltemp-earl-deem} that is known to perform well on such functions. Parallel tempering is a Markov chain Monte Carlo (MCMC) technique in which we run a parallel set of MCMC simulations, distributed across a range of temperatures. Each chain evolves by sampling at the specified temperature, but can probabilistically move between more global and more local exploration by exchanging states with a chain at another temperature, higher or lower (Figure~\ref{fig4}; Supporting Information S3). Each MCMC thus takes a random walk across temperatures that simulates a tempering process, which improves their ability to escape being trapped in local optima. 

To evaluate the null hypothesis that parasites and free-living species follow similar feeding rules, we measured the quality of the estimated model, using three measures described below, as we added progressively more parasite information to a free-living food web. In this sequence, there are three versions of the food web: (i) {\it FlensFree} or $G_F$, containing all free-living species $V_F$ and their predation links $E_{F \rightarrow F}$; (ii) {\it FlensPar} of $G_{FP}$, containing all free-living species $V_F$ and parasites $V_P$, the edges of $G_F$ as well as parasite-host links $E_{P\rightarrow F}$, predation among parasites $E_{P\rightarrow P}$, and predation on parasites $E_{F \rightarrow P}$; and (iii) {\it FlensParCon} or $G_{FPC}$, containing the same nodes and links as $G_{FP}$ as well as concomitant links $E_C$, i.e., $G_{FPC}$ includes predation and concomitant predation on parasites $E_{F \rightarrow^c P}$. The observed interaction types are made explicit in Figure~\ref{fig2}.

If parasites and free-living species follow distinct sets of rules, the quality of the model will decline as we require it to fit increasingly distinct types of feeding patterns, i.e., from \flensfree to \flenspar to $G_{FPC}$. Our measures of model quality are three-fold: (i) the goodness-of-fit for the model, formalized as an AUC statistic (see Supporting Information S2) on the observed predation links, which quantifies the ability of the model to correctly distinguish between observed predation links and observed non-feeding pairs; (ii) the fitted model’s ability to generate synthetic food webs with statistically similar structure to the empirical data via standard network measures; and (iii) the out-of-sample prediction accuracy for missing links, formalized as an AUC statistic in which we remove a subset of links from the empirical web and measure the model’s ability to accurately identify which edges were removed. We point out that these measures are comparable across food webs with different numbers of species $S$ and links $L$.

Applying this method to the Flensburg Fjord food web, we found that the PNM fits the data well, consistently yielding high AUC scores for in-sample goodness-of-fit  for each version of the web (Table~\ref{table1}). This indicates that the model is able to find an underlying niche structure that differentiates the probabilities of observed consumer links from the probabilities of pairs of species that do not consume each other. However, for the two food web versions including parasites, the AUC is much lower on parasite-host links, $E_{P\rightarrow F}$. On the other hand, goodness of fit is highest on $E_{F \rightarrow^c P}$, indicating that concomitant links are well captured by the PNM. We also find that trophic interactions among parasites $E_{P\rightarrow P}$ are fit as well as other types of predation links. Finally, the free-living species web \flensfree is slightly better fit than the food webs including parasites, measured across all observed edge types, which is consistent with scaling effects observed in Dunne et al.~\cite{dunne-2013-parafw}.

\begin{table*}
\caption{{\bf PNM goodness of fit.} Goodness-of-fit (AUC) statistics for models fitted to the entire Flensburg Fjord food web but evaluated on different subsets of predation links. These high values indicate the model can simultaneously explain predation among and between both free-living species and parasites. Parenthetical values indicate the standard error calculated on the optima found from 100 independent runs of the algorithm. 
\label{table1}}
\begin{ruledtabular}
\begin{tabular}{l|ccc}
& \multicolumn{3}{c}{\bf Goodness of fit AUC} \\\hline
{\bf Trophic interaction links type} & $\mathbf{G_F}$ & $\mathbf{G_{FP}}$ & $\mathbf{G_{FPC}}$   \\\hline
Among free-living $E_{F\rightarrow F}$ 	
				& 0.949 (0.009)	& 0.918 (0.017)	& 0.902 (0.018) \\
Among parasites $E_{P\rightarrow P}$ 	
				& ---	& 0.910 (0.044)	& 0.915 (0.034)	\\
From parasites to free-living $E_{P\rightarrow P}$ 	
				& ---	& 0.851 (0.040)	& 0.826 (0.039)	\\
From parasites to free-living $E_{P\rightarrow P}$ 	
				& ---	& 0.937 (0.024)	& ---	\\
From parasites to free-living $E_{P\rightarrow P}$ 	
				&  ---	& ---			& 0.974 (0.006)	\\
All links $E$ 	& 0.949 (0.009)	& 0.911 (0.013)	& 0.920 (0.009) \\

\end{tabular}
\end{ruledtabular}
\end{table*}

We compared statistical network properties of the empirical data to synthetic food webs drawn from the fitted PNM models. We found close agreement between the synthetic webs and the empirical ones for standard measures of network properties such as connectance and clustering coefficient; the average shortest path length between species in the data was slightly longer than in the resampled networks (Table~\ref{table2}). Overall, this suggests that the probabilistic niche model generates an ensemble of networks that are structurally similar to the original data. This was true across $G_F$, $G_{FP}$, and $G_{FPC}$, suggesting there is no scale dependence, nor sensitivity to parasites, on the ability of the model to fit and generate structurally similar networks.

\begin{table*}
\caption{{\bf Network properties of the original and resampled webs.} Network statistics of the original (observed) data and resampled networks from model fit to the data, using maximum likelihood estimates.
\label{table2}}
\begin{ruledtabular}
\begin{tabular}{l|cc|cc|cc}
{\bf Network measure} & $\mathbf{G_F}$  & $\mathbf{G_F}$ {\bf resampled} & $\mathbf{G_{FP}}$  & $\mathbf{G_{FP}}$ {\bf resampled} & $\mathbf{G_{FPC}}$   & $\mathbf{G_{FPC}}$  {\bf resampled} 
\\\hline
Number of species $S$	 	& 56 & 56 & 109 & 109 & 109 & 109 \\
Number of links $L$		 	& 358 & 369.4 (14.9) & 846 & 886.28 (34.0) & 1252 & 1293 (35.2) \\
Mean degree $\langle k \rangle$ 	& 6.393 & 6.597 (0.267) & 7.762 & 8.131 (0.031) & 11.486 & 11.87 (0.323) \\
Connectance $L/S^2$ 		& 0.114 & 0.118 (0.005) & 0.071 & 0.075 (0.003) & 0.105 & 0.109 (0.003) \\
Clustering coefficient 		& 0.227 & 0.260 (0.034) & 0.232 & 0.231 (0.025) & 0.355 & 0.377 (0.025) \\
Mean shortest path length 	& 1.966 & 1.863 (0.032) & 2.222 & 2.002 (0.030) & 1.972 & 1.821 (0.013) \\
\end{tabular}
\end{ruledtabular}
\end{table*}

Comparing predictions of presence or absence of missing links, we applied the link prediction goodness-of-fit test to different subsets of each of the three webs. We conducted a strong test of the ability of a single underlying niche structure to model both free-living and parasitic feeding links. We simulated an out-of-sample test by removing a uniformly random 10\% of observed links (i) from among free-living species $E_{F \rightarrow F}$, (ii) from free-living species to parasites $E_{F \rightarrow P}$ (in $G_{FP}$) or $E_{F \rightarrow^c P}$ (in $G_{FPC}$), (iii) from parasites to free-living species $E_{P\rightarrow F}$, or (iv) from among parasites $E_{P\rightarrow P}$, and fitting the model to the reduced food web (for $G_F$, this can only be done on $E_{F\rightarrow F}$). We then measured its ability to correctly place higher probabilities on the missing edges than on all other non-predation links in the graph~\cite{clauset08-hrg}. The AUC scores for each of these tests quantifies the amount of information the niche structure of the remaining links contains about the missing links of a given type. We found that differences across the different subwebs are not significant, which suggests the extra information (parasites; concomitant links) is not violating the model’s assumptions (Table~\ref{table3}). 

\begin{table*}
\caption{{\bf Link prediction with links withheld by type.} 
Link prediction on subwebs using AUC. For each web $G_F$, $G_{FP}$, and $G_{FPC}$, ten percent of links are dropped from each subweb (listed in left column). The AUC is then calculated over the true non-links and false non-links (``missing links'') of the original web. Parenthetical values indicate the standard error calculated on the optima found from 100 independent runs of the algorithm.
\label{table3}}
\begin{ruledtabular}
\begin{tabular}{l|ccc}
 & \multicolumn{3}{c}{\bf Missing links AUC }
\\\hline
{\bf Missing trophic interaction links type} & $\mathbf{G_F}$ & $\mathbf{G_{FP}}$ & $\mathbf{G_{FPC}}$   \\\hline
Among free-living species $E_{F\rightarrow F}$	 
				& 0.844  (0.053) & 0.849 (0.032) & 0.834 (0.039)  \\
Among parasites $E_{P\rightarrow P}$	 
				& --- 	& 0.842 (0.031) & 0.871 (0.032)  \\
From parasites to free-living $E_{P\rightarrow F}$	 
				& --- 	& 0.854 (0.032) & 0.864 (0.034)  \\
From free-living to parasites $E_{F\rightarrow P}$ excl.\ $E_C$	 
				& --- 	& 0.835 (0.034) & ---  \\
From free-living to parasites $E_{F\rightarrow^c P}$ incl.\ $E_C$	 				& --- 	& --- 	& 0.868 (0.038)  \\
All links $E$	& 0.844  (0.053) & 0.813 (0.023) & 0.840 (0.017)  \\
\end{tabular}
\end{ruledtabular}
\end{table*}

Finally, we checked whether the model has overfitted the data, as a measure of the robustness of our results (Table~\ref{table4}). We removed a uniformly random 10\% of observed links from each web and trained the model on the reduced food web. We then compared the probabilities of all true (observed) links to the probabilities of the true non-links. We also broke down the comparison of true links and non-links by link-type. If the model were overly sensitive to the missing links, then we expect the model to perform less well, predicting the missing links poorly. We found that the noisy in-sample AUC scores are comparable to the scores on the fully observed data (Table~\ref{table1}), which suggests that the model is not overly sensitive to noise, i.e., not overfitting the data. As in Table~\ref{table1}, parasite-host links $E_{P\rightarrow F}$ are least well fit by the model, but trophic interactions among parasites are comparably well fit to other predation link types; concomitant links are again well explained by the model.

\begin{table*}
\caption{{\bf Robustness of the data and subsets, model fitted to noisy under-sampled data.} 
Robustness on the data and subsets, with the model fitted with 10\% of the observed links withheld. Parenthetical values indicate the standard error calculated on the optima found from 100 independent runs of the algorithm.
\label{table4}}
\begin{ruledtabular}
\begin{tabular}{l|ccc}
 & \multicolumn{3}{c}{\bf Robustness AUC} \\\hline
{\bf Trophic interaction links type} & $\mathbf{G_F}$ & $\mathbf{G_{FP}}$ & $\mathbf{G_{FPC}}$    \\\hline
Among free-living species $E_{F\rightarrow F}$	 
				& 0.934  (0.014) & 0.907 (0.022) & 0.895 (0.019)  \\
Among parasites $E_{P\rightarrow P}$	 
				& --- 	& 0.924 (0.042) & 0.905 (0.037)  \\
From parasites to free-living $E_{P\rightarrow F}$	 
				& --- 	& 0.840 (0.042) & 0.805 (0.028)  \\
From free-living to parasites $E_{F\rightarrow P}$ excl.\ $E_C$	 
				& --- 	& 0.936 (0.030) & ---  \\
From free-living to parasites $E_{F\rightarrow^c P}$ incl.\ $E_C$	 				& --- 	& --- 	& 0.974 (0.007)  \\
All links $E$	& 0.934  (0.014) & 0.900 (0.013) & 0.912 (0.008)  \\
\end{tabular}
\end{ruledtabular}
\end{table*}

As a number of previously hypothesized models of niche structure are special cases of the PNM, we examined the inferred parameter values for the corresponding patterns that would indicate support for two alternate models, the cascade model and the inverse niche model~\cite{inverse-niche-model,cohen-newman-cascade,stouffer2005cascade}. The cascade model requires that consumers only feed on those below them in the niche space, whereas the inverse niche model follows a niche model on predation among free-living species as in Ref.~\cite{williams-martinez-niche-2000}, and parasites feeding on free-living hosts above them in the niche space, with feeding range width decreasing with higher niche position for parasites. We consider 100 local optima for each web and search for these properties within each optimum and on average.  In no case did every species in the model have an inferred $n_i \ge c_i$, i.e., a strict cascade. On average, and for both free-living species and parasites, there is no statistically significant direction of feeding: the average $n_i - c_i$ isn't statistically different than zero for any type of species. In no case did every species follow the inverse niche model, where free-living species follow the cascade model ($n_i \ge c_i$) and parasites follow an inverse cascade ($n_i \le c_i$) (see Supporting Information S4 and Table S4). Contrary to the inverse niche assumption, niche position and feeding range width are uncorrelated for parasites. Feeding niches were also not continuous~\cite{PNM2010,Zook2010}. Looking past those specific models, all parameters $n_i$, $c_i$, and $r_i$ were distributed significantly differently for free-living species than for parasites (KS test, $p< 0.001$; Supporting Information S4, Table S5).

\section{Discussion}

We found that there is little evidence that there is a structural distinction between parasites versus free-living species with respect to the model’s ability to learn the structure of predation in a real food web. The PNM accurately represented predation among free-living species, predation on parasites, concomitant predation, and predation among parasites. Furthermore, we found that predation is well explained by the niche model, regardless of consumer or resource type. The similarity of predation on parasites to predation among free-living species sheds light on the poorly-understood role of predation on parasites~\cite{johnson-parasites-prey}. The PNM was able to successfully model predation even without additional allowances for secondary niches or separation by life stage~\cite{johnson-parasites-prey,inverse-niche-model,dunne-2013-parafw}; other work suggests that separation by life stage would not improve the fit of the PNM~\cite{preston2014complex}. Conversely, parasite-host interactions were less well described by the PNM. In this food web, parasites play similar roles to free-living predators, i.e., predation is predation, regardless of context or body size, but parasitism is a structurally distinct trophic strategy. 

Our results showed that parasites occupy a broad range of niches, interspersed among the niches occupied by free-living species, but parasite niches are distributed differently than free-living niches. Despite this difference, separating parasites and free-living species may not be a necessary or meaningful distinction in describing the structure of predatory interactions. Other traits, such as niche width or relative abundance, may prove to be more useful features for modeling heterogeneities in food web link structure. 

The general nature of the PNM also allows us to test for the signature of specific structuring mechanisms that represent alternative models of food web structure. For instance, the cascade model, the simplest and earliest food web model~\cite{cohen-newman-cascade}, embodied the notions that taxa feed with a fixed probability on species with lower niche values, and that their niche is non-contiguous.  Hierarchical feeding is at the heart of subsequent niche and related models, although in a relaxed form~\cite{williams-martinez-niche-2000,cattin2004phylogenetic,stouffer2005cascade}, and the niche model further embodies contiguous feeding niches. Another example that is a special case of the PNM is the recent inverse niche model of free-living predation and parasitism on free-living hosts. The inverse niche model keeps a relaxed, contiguous feeding hierarchy for free-living species but reverses its direction for parasites, which feed on free-living taxa with higher niche values~\cite{inverse-niche-model}. In our analysis of the Flensburg food web, we did not find evidence for the cascade or the inverse niche models using the PNM. The inverse niche model did not model predation on parasites or predation among parasites, $E_{F\rightarrow P}$, $E_{F\rightarrow^c P}$, or $E_{P\rightarrow P}$, so there is no explicit comparison possible for those interactions. However, the poor fit of links from parasites to free-living species $E_{P\rightarrow F}$ suggests that an alternate mechanism, perhaps similar to the inverse niche model, may be necessary to explain such parasite-host connections.

There has been disagreement about whether concomitant predation links should be included in food web data, in part because they represent a secondary form of trophic interaction compared to classic predation or parasitism (Figure~\ref{fig2})~\cite{Marcogliese-parasiteplea,para-missing-links,parasites-dominate-fw}. These secondary links embed information about trophic intimacy between parasites and hosts, and it is currently unclear what structural or functional roles these links play in food webs~\cite{johnson-parasites-prey,dunne-2013-parafw}. Here, we found that concomitant links did not obscure the underlying niche structure of either free-living species or parasites, and including them led to no significant decrease in model fit. In fact, predation on parasites was easier to predict when concomitant links were included. Concomitant links were naturally represented by the PNM and appeared to follow similar patterns to other types of predation on parasites. 

Previous work found that food webs including concomitant links deviated in motif frequencies~\cite{dunne-2013-parafw,cirtwill2015concomitant}. However, motifs and niches describe the network at fundamentally different levels, and so there is no conflict between such observations and our results. Concomitant links naturally close triangles and can create bidirectional links between parasites whose hosts also consume the parasites as concomitant prey. These bidirectional links are relevant as a mode of trophic parasite transmission and infection between free-living hosts. At the motif level, these triangles and bidirectional links obscure motif distributions, but here, they reinforce the niche structure and increase predictability (Supporting Information S5). 
Under the PNM, the global properties of the network, including our network measures and the distinction between links and non-links, are preserved when concomitant links are included. Dunne et al.~\cite{dunne-2013-parafw} found that the roles of parasites as consumers were different from free-living species. We found that this difference splits by type of resource: specifically, when the resources are free-living, we find that the PNM represents these parasitic trophic interactions less effectively. When the resources are other parasites, the PNM represents these links easily, and as easily as when the consumer is free-living. Dunne et al.\ also found that parasites have more complex trophic niches, which reduces the goodness of fit for the PNM on predicting parasite consumer links. Complex trophic niches corroborate the lessened predictive ability of the PNM on parasite-host links. 

Generative models, such as the PNM, are a sophisticated tool for investigating the structure of food webs, including whether different types of taxa and interactions follow distinct connectivity patterns. We united these techniques and applied them to a single food web. By applying these methods to a wide variety of food webs, one can assess the generality of these results. Applying such techniques to a broader range of data and to other types of trophic interactions and species will help characterize structural differences, generate novel ecological hypotheses, and support the iterative development and testing of ecological models and theory for parasites and other previously underrepresented taxa and interactions~\cite{kefi2012merging,britton-2013,dunne-2013-parafw,thieltges-paraprey-2013}. Broadly, these methods provide a principled framework to detect heterogeneities in the roles of nodes and links in empirical network data.

\section{Methods}

\subsection{Data}
We use the Flensburg Fjord food web data~\cite{flensburg-fjord-data} to demonstrate our methods. We consider three nested subsets of the data: $G_{F}$, $G_{FP}$, and \flenscon (Figure~\ref{fig1}). We define the species set $V_F$ as all free-living and basal taxa, and $V_P$ as only parasites. We follow existing naming conventions to distinguish these data sets, which are constructed as follows:
\begin{itemize}
\item FlensFree ($G_{F}$) contains only links between free-living and basal taxa.
\subitem \flensfree includes taxa $V_F$ and predation links $E_{F\rightarrow F}$
\item FlensPar ($G_{FP}$)
\subitem \flenspar includes taxa $V_F$ and $V_P$ and predation links $E_{F\rightarrow F}$; predation on parasites, excluding concomitant links, $E_{F\rightarrow P}$; parasite-host links $E_{P\rightarrow F}$; and predation among parasites $E_{P \rightarrow P}$ 
\item FlensParCon ($G_{FPC}$)
\subitem \flenscon includes taxa $V_F$ and $V_P$ and predation links $E_{F\rightarrow F}$; predation on parasites, including concomitant links, $E_{F\rightarrow^c P}$; parasite-host links $E_{P\rightarrow F}$; and predation among parasites $E_{P \rightarrow P}$ 
\end{itemize}
We consider four subsets of the webs in our analyses, corresponding to the four quadrants of Figure~\ref{fig2}: links (i) among free-living species $V_F \times V_F$; (ii) from free-living species to parasites $V_F \times V_P$, possibly including concomitant links; (iii) from parasites to free-living species $V_P \times V_F$; and (iv) among parasites $V_P \times V_P$, representing the sets of potential consumer-resource relationships. The elements of the subgraph of $V_F \times V_P$ will vary dependent on the inclusion of concomitant links, either iwth edges $E_{F\rightarrow P}$ or $E_{F \rightarrow^c P}$. Due to trophic aggregation, the set of free-living species $V_F$ in $G_{FP}$ and $G_{FPC}$ is not equivalent to the set of free-living species in $G_F$.

\subsection{Probabilistic Niche Model}

The probabilistic niche model (PNM) of Williams et al.~\cite{PNM2010} and Williams and Purves~\cite{PNM2011} is a probabilistic construction of the niche model for food webs~\cite{williams-martinez-niche-2000} that creates quasi-interval webs~\cite{Zook2010}. For a food web of $S$ species, each species $i$ that resides in an ecological niche located at $n_i$ in the underlying one-dimensional niche space. Each species consumes other species in the food web with probability given from a species $i$'s feeding distribution with center at $i$'s ideal feeding position $c_i$ and variance $r_i$, corresponding to $i$'s feeding range (Figure~\ref{fig3}). We express the full vector of PNM parameters as $\theta = \{ \alpha, n_1, \ldots, n_S, c_1, \ldots, c_S, r_1, \ldots, r_S, e \}$.

The probability of species $i$ consuming species $j$ is given by:
\begin{equation}
\Pr(i \rightarrow j | \theta) = \alpha \exp \left(-\left|\frac{n_j - c_i}{r_i / 2} \right| \right)
\end{equation}
where $\alpha$ is an uncertainty parameter traditionally set to 0.9999 and $e$ can take values other than 2 for a broader range of distributions~\cite{PNM2010}. When $\alpha$ is allowed to be a free parameter, we find values near 1, so we fix this parameter in practice. We allow $e$ to be a free parameter, and find that its value decreases as we introduce parasite nodes and concomitant links (Table S3).

\subsection{Evaluating model performance}

The log-likelihood for the food web data $G$ given the parameters $\theta$ is defined as:
\begin{equation}
\mathcal{L}(G | \theta) = \sum_{i=1}^S \sum_{j=1}^S  \ln\left\{  \begin{array}{l l}
\Pr(i \rightarrow j | \theta) & \text{ if }G(i,j) = 1 \\
1 - \Pr(i \rightarrow j | \theta) & \text{ otherwise }
\end{array} \right. 
\end{equation}
for the PNM~\cite{PNM2010}. Deviations of the data from the predictions made by the model can then be observed, either as non-edges with predicted high probability ($G(i,j) = 0,\ \Pr(i\rightarrow j | \theta)$) or observed edges predicted with low probability ($G(i,j) = 1,\ \Pr(i\rightarrow j|\theta)$). Ranking all edges by predicted presence $Pr(i \rightarrow j|\theta)$ and comparing such a ranking to the observed $G(i,j)$ then describes the goodness of fit of the PNM. We calculate the AUC $A$ on the ranked probabilities, $x_i$ and $y_j$ over sets of size $S_1$ and $S_2$ as 
\begin{equation}
A = \frac{\sum_{i=1}^{S_1} \sum_{j=1}^{S_2} \mathbf{1}_{x_i > y_j}}{S_1 S_2}.
\end{equation}
We evaluate the performance of the model using AUC, or the area under the receiver operating characteristic curve. The AUC measures the separation of the distributions of probabilities predicting true links from true non-links. Intuitively, the AUC is the probability that given a true link and a true non-link, we rank the true link higher than the true non-link. AUC has high natural variance on sets of disparate sizes, as it effectively oversamples the smaller set to calculate that probability. Sets such as links (of size $L$) and non-links (of size $S^2 -L$) will be of disparate sizes when connectance is low, typical of food webs ($L/{S^2} \ll 1$).

\subsection{Optimization for the PNM with parallel tempering}

We use parallel tempering to find optima of the maximum likelihood parameters. Parallel tempering, also known as replica exchange MCMC (Markov chain Monte Carlo), is an efficient and easily parallelizable optimization technique from statistical physics~\cite{paralleltemp-earl-deem,newman-barkema-MCMC}.

In parallel tempering, $Q$ replicas of the system (likelihood space) are explored using MCMC, under the Metropolis-Hastings algorithm. The prelicas are taken over a range of mixing temperatures $T_1, \ldots, T_Q$, and are run in parallel. After every $\tau$ steps of the chain, a pair of replicas at adjacent temperatures $T_k, T_{k+1}$ is allowed to switch location in the psace with probability based on their relative likelihood and temperatures (Figure~\ref{fig3}). The replicas switch with probability 
\begin{equation}
p = \min\{1, \exp(\frac{1}{T_i} - \frac{1}{T_j})(\mathcal{L}_i - \mathcal{L}_j) \}.
\end{equation}

Parallel tempering is a general MCMC method and meets the detailed balance condition. By combining high temperature (fast-mixing) and low temperature (slow-mixing, or locally hill-climbing) chains, parallel tempering allows us to more quickly survey the likelihood space and explore more diverse local optima~\cite{paralleltemp-earl-deem}. See Supporting Information S3 for more details and guidelines.

\begin{acknowledgments}
We thank Rich Williams and Daniel Stouffer for helpful conversations. This work was supported in part by Grant \#FA9550- 12-1-0432 from the U.S. Air Force Office of Scientific Research (AFOSR) and the Defense Advanced Research Projects Agency (DARPA) (AZJ, CM, AC); the NSF GRFP award DGE 1144083 (AZJ); NSF grant IIS-1452718 (AC), and the Santa Fe Institute. 
\end{acknowledgments}

%
\end{bibunit}

\newpage
\begin{bibunit}
\section*{Supplementary information}

\section*{S1. Data.}

We apply our methods to the Flensburg Fjord food web of consumer-resource interactions~\cite{flensburg-fjord-data}. Taxa with complex life cycles are aggregated over life stages, and taxa are trophically aggregated as in Ref.~\cite{flensburg-fjord-data}. We defer to the definitions of parasites as defined by the data collectors. We consider three nested subsets of the data: $G_{F},\ G_{FP},\ G_{FPC}$ (Figure 1, main text). We follow existing naming conventions to distinguish these data sets, which are constructed as follows:
\begin{itemize}
\item \textbf{FlensFree} (${G_F}$) contains only links between free-living and basal taxa.
\item \textbf{FlensPar} (${G_{FP}}$) contains parasite, free-living, and basal taxa, but excludes concomitant predation. The web still includes links from free-living species to parasites, but limited only to predation on parasites.
\item \textbf{FlensParCon} (${G_{FPC}}$) contains parasite, free-living, and basal taxa, and includes concomitant predation links. 
\end{itemize}
We use the following convention for arrows in all figures and notation:
\begin{equation*}
\text{Consumer}\rightarrow\text{Resource}
\end{equation*}
that is, in the direction of ``Consumer \textit{feeds on} Resource'' and in the opposite direction of the energy flow. 

We  consider four subgraphs in our analyses (Figure 2, main text):
\begin{itemize}
\item \textit{Free consumers, Free resources}: The subgraph representing predation among free-living species. Vertices are free-living taxa $V_{F}$, predation links are $E_{F\rightarrow F} \subseteq V_{F} \times V_{F}$.
\item \textit{Parasite consumers, Parasite resources}: The subgraph representing all trophic interaction among parasites. Here only including intraguild trophic interaction representing competition among parasites, but could potentially include hyperparasitism. Vertices $V_{P}$, edges $E_{P\rightarrow P} \subseteq V_{P}\times V_{P}$.
\item \textit{Free consumers, Parasite resources}: The subgraph representing direct predation on parasites (links $E_{F\rightarrow P}$) and potentially concomitant predation (links $E_c$). When taken as a subgraph of $G_{FP}$, the subgraph has edges $E_{F\rightarrow P} \subseteq V_{F} \times V_{P}$. As a subgraph of $G_{FPC}$, the subgraph has edges $E_{F\rightarrow^{c}P} = E_{F\rightarrow P} \cup E_{c} \subseteq V_{F} \times V_{P}$. 
\item  \textit{Parasite consumers, Free resources}: The subgraph representing parasite-host interactions, for free-living hosts. Vertices are $V_{P} \cup V_{F}$, and edges are only the parasite-host links, $E_{P\rightarrow F} \subseteq V_{P} \times V_{F}$.
\end{itemize}
When we refer to a subgraph, we specify from which web we take a subset. The subgraphs involving parasites are, by definition, only taken from the FlensPar ($G_{FP}$) and FlensParCon ($G_{FPC}$) webs.

The set of consumers and resources labeled \textit{Free-living} includes all free-living and basal taxa, and \textit{Parasites} includes only parasites. We occasionally define the subgraphs by their link sets, where, e.g., $E_{{F\rightarrow P}} \subseteq E_{F\rightarrow^c P} \subseteq E$ and $E_{{F\rightarrow F}} \subseteq V_{F} \times V_{F}$. It is worthwhile to note that the set of free-living species in FlensPar and FlensParCon is not equivalent to FlensFree, due to trophic aggregation.

The number of species and links, and the number by type, are shown in Table S1. Basic properties of the webs, including connectance, are given in Table S2. The in- and out-degree distributions are shown for the food webs in Figure S1. These degree distributions are broken down by type (free-living and parasite) in Figure S2.

\begin{table*}[htbp]
\begin{centering}
\begin{small}
\begin{ruledtabular}
\begin{tabular}{rr | cccc}
&&  $\mathbf{G_{F}}$ &  $\mathbf{G_{FP}}$ &  $\mathbf{G_{FPC}}$  \\
\hline
& Species ($S$) & 56 & 109 & 109 \\
& Links ($L$) & 358 & 846 & 1252 \\
\hline
& Free-living species $V_{F}$  & 56 & 68 & 68 \\
& Parasite species $V_{P}$ & --- & 41 & 41 \\
\hline
& Link type $E_{{F\rightarrow F}}$ & 358 & 520 & 520 \\
& $E_{F\rightarrow P}$ & --- & 64 & 64 \\
& $E_{c}$ & --- & --- & 406  \\
& $E_{F\rightarrow^{c}P}$ & --- & --- & 470 \\
& $E_{P\rightarrow F}$ & --- & 227 & 227 \\
& $E_{P\rightarrow P}$ & --- & 35 & 35 \\
\hline
& Basal species ($V_{\text{basal}} \subseteq V_{F}$) 	& 6 & 6 & 6 	\\	
& Specialist free-living species (out-degree 1) 		& 2	& 4	& 4	\\
& Generalist free-living species (out-degree $>$ 1)	& 48	& 58	& 58	 \\
& Specialist parasite (out-degree 1)				& --- & 7	& 7	\\
& Generalist parasite (out-degree $>$ 1)			& --- & 34	& 34 	\\
\end{tabular}
\end{ruledtabular}
\end{small}
\captionsetup{labelformat=empty,justification=centering}
\caption{{\bf Table S1.} Number of species and link types in Flens trophic food webs.}
\label{table:popflens}
\end{centering}
\end{table*}
\begin{table*}[htbp]
\begin{centering}
\begin{small}
\begin{ruledtabular}
\begin{tabular}{r | ccc}
&  $\mathbf{G_{F}}$ &  $\mathbf{G_{FP}}$ &  $\mathbf{G_{FPC}}$\\ 
\hline
Species ($S$) & 56 & 109 & 109 \\
Links ($L$) & 358 & 846 & 1252\\
\hline
Connectance ($\frac{L}{S^{2}}$) & 0.114 & 0.071 & 0.105 \\
Directed clustering coefficient & 0.001 & 0.012 & 0.009 \\
Clustering Coefficient & 0.227 & 0.232 & 0.355 \\
Reciprocity & 0.006 & 0.006 & 0.102 \\
Mean shortest path length & 0.314 & 1.166 & 1.446 \\ 
\hline
Mean in-degree $<k_{in}>$ (std. dev) & 6.39 (6.01) & 7.76 (7.51) & 11.49 (7.69)\\
Mean out-degree $<k_{out}>$ (std. dev) & 6.39 (6.15) & 7.76 (7.86) & 11.49 (14.71)\\
Maximum in-degree $k_{in}$		& 24 & 31 & 31 \\
Maximum out-degree $k_{out}$		& 29 & 41 & 77\\
Number of self-loops				& 2 & 2 & 2 \\
Correlation($k_{in}, k_{out}$) & -0.373 & 0.170 & 0.090 \\
\end{tabular}
\end{ruledtabular}
\end{small}
\captionsetup{labelformat=empty,justification=centering}
\caption{{\bf Table S2.} Descriptive statistics for Flens trophic food webs.}
\label{table:descflens}
\end{centering}
\end{table*}
\begin{figure*}[htbp]
\centering
\includegraphics[width=7in]{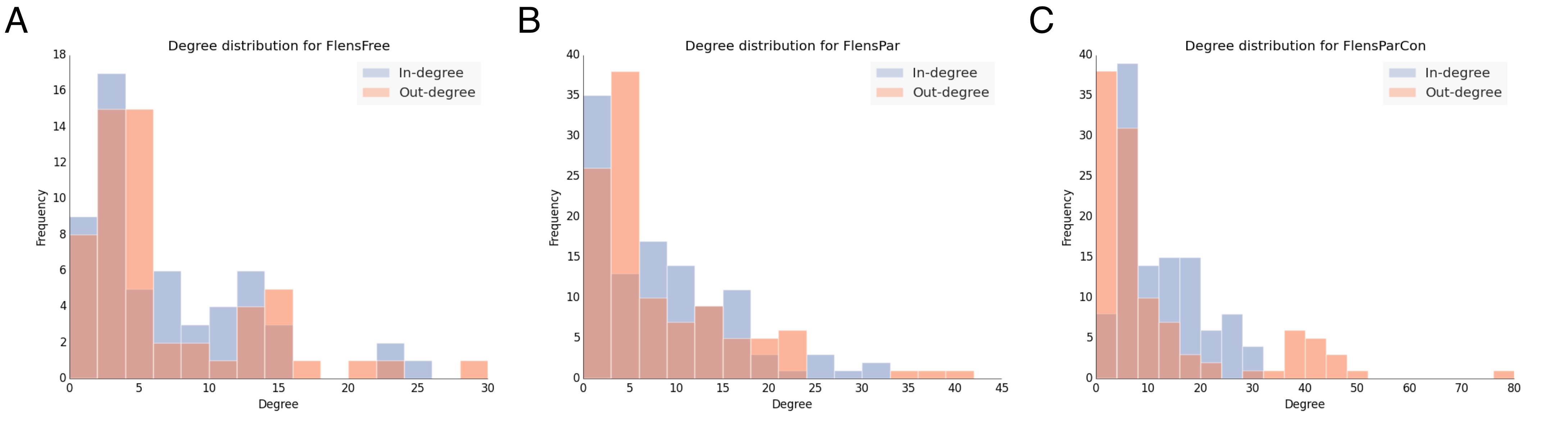} 
\captionsetup{labelformat=empty,justification=centering}
\caption{{\bf Figure S1.} In- and out-degree distributions for $G_{F}$ (FlensFree), $G_{FP}$ (FlensPar), $G_{FPC}$ (FlensParCon), parts A, B, and C, respectively.}\label{fig:degreeflens}
\end{figure*}
\begin{figure*}[htbp]
\centering
\includegraphics{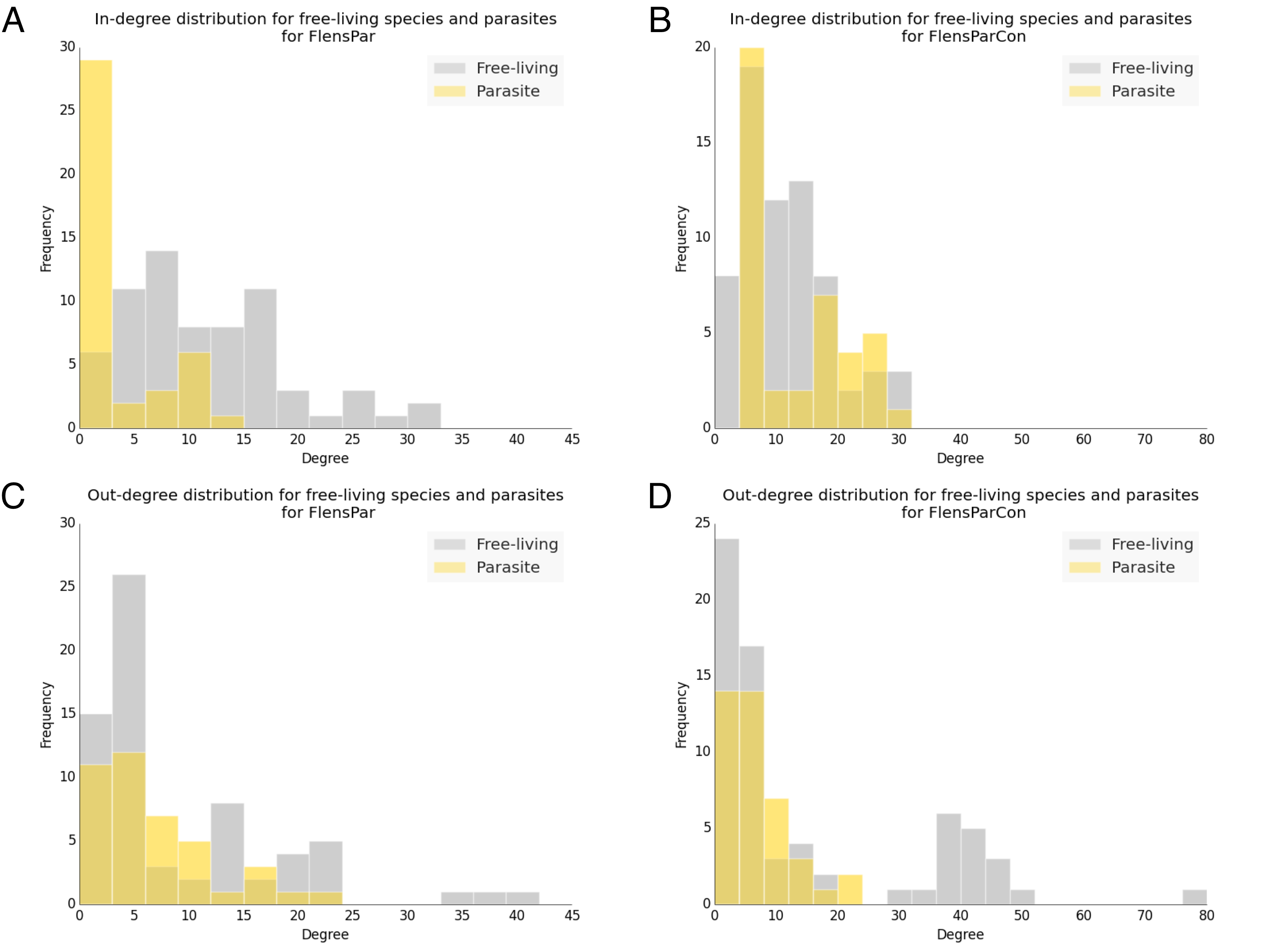} 
\captionsetup{labelformat=empty,justification=centering}
\caption{{\bf Figure S2.} In-degree distributions for free-living species and parasites in $G_{FP}$ (FlensPar) and $G_{FPC}$ (FlensParCon) in parts A and B, respectively. Parts C and D, out-degree distributions for free-living species and parasites in $G_{FP}$ (FlensPar) and $G_{FPC}$ (FlensParCon)}.\label{fig:degreefreepar}
\end{figure*}
%

\section*{S2. Probabilistic Niche Model definition and evaluation.}

\subsection*{S2.1 The Probabilistic Niche Model}
For a food web with $S$ species, we assume there exists some $d$-dimensional latent space, and we assume that each species resides in some location in that space~\cite{PNM2010,PNM2011}. This location corresponds to their ecological niche. We assume that trophic species should occupy the same niche and as such we do not lose any information about the web by grouping them together. In the original niche model, described in Ref.~\cite{williams-martinez-niche-2000}, a species only consumes other species contained by a particular contiguous region in niche space. In the probabilistic niche model, we loosen this assumption to let these feeding relationships be defined probabilistically. Rather than a deterministic and contiguous region, we assume that each species has a distribution over the niche space corresponding to their most likely consumption of species in the space. For each species, this feeding distribution is allowed to vary. This distribution is described with two parameters, the center, corresponding to their preferred feeding position, and the width. All possible feeding relationships to have a nonzero probability (although possibly very close to zero).

The full vector of parameters is given by $\theta = \{\alpha, n_1, \ldots, n_S, c_1, \ldots, c_S, r_1, \ldots, r_S, e \}$. For each species $i \in \{1, 2, \ldots, S\}$, we define three parameters: $n_{i},\ c_{i}$, and $r_{i}$. The position of species $i$ in the niche space is given by $n_{i}$, and species $i$ consumes other species $j$ with probability kernel (feeding distribution) centered at $c_{i}$ with variance $r_{i}$. Then the probability of species $i$ consuming species $j$ is:
\begin{equation}\label{si-pnm-edgeprob}
\Pr(i \rightarrow j|\theta) = \alpha \exp\left( - \left|\frac{n_j - c_i}{r_i/2} \right|^e \right).
\end{equation}
Here, $\alpha$ is an uncertainty parameter traditionally set to $0.9999$, which dampens the probability of observing each relationship. When $\alpha$ is allowed to be a free parameter, we find values near 1, so we fix this parameter in practice; this result was also found by Williams et al.~\cite{PNM2010}. We define the niche space to be the $[0,1]$ interval, and allow species to be embedded within that space. That is, we constrain $0 \le n_i \le 1$ and $0 \le c_i \le 1$ for all species $i$. The parameters describing the width of species' feeding ranges $r_i$ and the global parameter describing the curvature of the distribution $e$ are only required to be positive, but take on a potentially broader range of values. The distribution-shape parameter $e$ creates a traditional Gaussian distribution when $e=2$, but we allow it to take other values to allow for a broader range of distributions~\cite{PNM2010}. We allow $e$ to be a free parameter. We find that the parameter $e$  decreases, thus creating a more peaked distribution, as we introduce parasite nodes and concomitant links (Table S3). 

\begin{table}[tbhp]
\centering
\begin{tabular}{c|c}
\hline\hline
{\bf \ \ \ Data\ \ \ } & {\ \ \bf Parameter} $e$ {\bf mean (std.\ err)\ \ }\\\hline
$G_{F}$ 		& 0.874 (0.289) \\
$G_{FP}$ 		& 0.516 (0.111) \\
$G_{FPC}$ 	& 0.500 (0.010) \\\hline
\end{tabular}
\captionsetup{labelformat=empty,justification=centering}
\caption{{\bf Table S3} Kernel decay parameter $e$ for 100 independent runs of the algorithm}
\label{tab:kernele}
\end{table}

The feeding range width parameter $r_i$ roughly corresponds to the breadth of species $i$'s diet. A large value would correspond to more generalist species, who are likely to eat from a wider range of niches. However, this relationship is not exact: for example, if many species all reside in the same niche, and a consumer consumes within that niche with high probability and small width, then a generalist $i$ can be captured by a small feeding range width parameter $r_i$.

\subsection*{S2.2 Evaluating the model performance}

Using the edge probabilities defined in Eqn.~\ref{si-pnm-edgeprob}, the log-likelihood for the full food web $G$ for the parameters $\theta$ is then: 
\begin{equation}\label{si-pnm-lik}
\mathcal{L}(G | \theta) = \sum_{i=1}^S \sum_{j=1}^S  \ln\left\{  \begin{array}{l l}
\Pr(i \rightarrow j | \theta) & \text{ if }G(i,j) = 1 \\
1 - \Pr(i \rightarrow j | \theta) & \text{ otherwise }.
\end{array} \right. 
\end{equation}
We use this equation as an objective function to fit the model, that is, to find the parameter values for the model given the data. This function will give low values when the parameters poorly describe the data (give low probability to true links and vice versa), and higher values when they better describe the data (give high probability to true links and vice versa). However, as this objective function is used to fit the model (parameters) to the data, this is not sufficient to argue whether or not the model itself is sensible for the data. Goodness of fit and link prediction tasks can help evaluate the model performance, and with these tests, we can argue that the PNM is (or isn't) a sensible choice for food web data, with or without parasites and concomitant links. 

In practice, there may be some links that are not observed, $G(i,j) = 0$ in the data, and yet have high probability $Pr(i \rightarrow j| \theta)$. This may be due to a poor choice of model or poor choice of parameters for the model, but ideally these links are assigned high probability due to some other mechanism. For example, if those links are ecologically likely, they may be true `missing links' that were unobserved in this data. This allows for discovery driven by the generative model itself. We may also create the analogous scenario with a link prediction task. In that task, we artificially change some observed edges $G(i,j) = 1$ to be `unobserved' with $G(i,j) = 0$. If we fit the model to that data, but find that those links are predicted with high probability, then this means the model may be of high quality. In particular, in the link prediction task, we test if the model assigns probability differently to true but artificially-withheld links from true non-links.

Similarly, some links may be given low probability $\Pr(i \rightarrow j|\theta)$ but will be observed, $G(i,j) = 1$. The goodness of fit tasks are designed to test how well separated in probability true links are from true non-links. If the model is failing to represent the data well, the model won't necessarily put high probabilities across all links and low probabilities across non-links. If the model is failing on a particular {\it type} of link, then looking at the the goodness of fit by link type can allow us to detect systematic failures of the model. This would suggest that the model may be appropriate to represent some types of links and not others. 

Detecting and quantifying these departures from the probabilistic niche model allows us to understand how the model can fail to represent the data, including whether or not the probabilistic niche model is appropriate for specific taxa, types (such as parasites), and types of relationships (such as parasite-host interactions). Uniting these tests allows us to more sensitively test when and how different models fail to describe observed ecological phenomena. 

Previous work has used the ``expected fraction of links predicted correctly'' measure $f_L$, the measure previously used to evaluate performance of the niche model~\cite{PNM2010, PNM2011, dunne-2013-parafw}. The measure takes the average of probabilities, assigned by the model, over each observed link, that is:
\begin{equation*}
f_L = \frac{1}{L}\sum_{i=1}^S \sum_{j=1}^S \Pr(i \rightarrow j|\theta) \text{G}(i,j) .
\end{equation*}
This is analogous to an accuracy score. This measure rewards models that assigns true links high probabilities, but with no penalty for assigning true non-links high probability as well. Ideally our model should assign both high probability to observed links and low probability to unobserved links in a way that \text{separates} observed and unobserved links. For $f_L$, this is mildly enforced by requiring a similar number of total expected links, but not systematically. 

Instead, we use the AUC to summarize these patterns in goodness or failure to fit the data. The AUC, the area under the receiver operating characteristic curve, is a direct, model-based measure for link prediction and goodness of fit. After we fit the model, we rank all edges by probability $\Pr(i \rightarrow j|\theta)$, and compare this ranking to the observed $G(i,j)$. The AUC summarizes this ranking. AUC measures how well separated the distributions of true links and true non-links will be; maximizing the likelihood of the model will push these distributions higher and lower, respectively. We focus on the detection of false negatives (missing links) in the link prediction and robustness tests. In general assume the data contain no false positives (incorrectly observed links). 

\subsubsection*{Measuring model performance using the AUC}

We evaluate the performance of the model using AUC. 
AUC, the area under the receiver operating characteristic (ROC) curve, is used to quantify the discriminatory power of a binary classifier. The statistical properties of the AUC are well-understood (e.g.,~\cite{largesampletechniques}). 
Intuitively, the AUC is the probability, given a true link and a true non-link, that we rank the true link higher than the true non-link. This metric has been used in other contexts specifically for link prediction~\cite{clauset08-hrg}, and here we use it for link prediction and goodness of fit. This measure explicitly lets us evaluate how we separate true observed links and true non-links. 

To calculate AUC, consider comparing two samples, Sample 1 of size $S_1$, and Sample 2 of size $S_2$. For goodness of fit tasks, we compare observed links $S_1 = L$ and observed non-links $S_2 = S^2 - L$. For link prediction tasks, we compare withheld but true links $S_1 = \#\{\text{links withheld}\} \le L$ from true non-links $S_2 = S^2 - L$. (For these tasks, we fit the PNM to the data, calculate the link probabilities, and then rank the links by their probability assigned by the model.) We compare the sum of ranks of Sample 1 to the sum of the top $S_{1}$ ranks, $S_{1}(S_{1}+1)/2$, such that:
\begin{equation}\label{auc-eqn}
\textrm{AUC} = \frac{\sum_{s \in \text{ Sample 1}}\text{rank}(s) - \frac{S_{1}(S_{1}+1)}{2}}
{S_{1}S_{2}}.
\end{equation}
As given here, AUC takes values between 0 and 1; values close to 0 or 1 reflect better discriminatory power. The AUC takes the value 0 for having all elements in Sample 1 ranked first and separate from all elements in Sample 2, and it takes the value 1 when all elements in Sample 1 are ranked last. For a random classifier, we would expect an AUC value of about 0.5, effectively a uniform random draw of ranks from $[1, 2, \ldots, S_{1}+S_{2}]$. In practice, we report the value of AUC to be $A =\max(\text{AUC}, 1-\text{AUC})$. Then we only report scores between 0.5 and 1. Good discriminatory power is reflected by higher AUC values, and $\text{AUC}=0.5$ corresponds to a random classifier, as before. Finally, we also make note of the distribution of AUC scores over different samples, as the natural variance of the AUC will be high in tests where the sample sizes are very different, e.g., where $S^2 - L \gg L$.

As we previously established, the AUC measure compares sampled probabilities of true links and true non-links. This means that the measure may perform differently in the low connectance ($L/S^2 \ll 1$) setting which is typical to food webs. In particular, the AUC will have high variance on sets of disparate sizes (such as sparse links, $L$ small, and many non-links, $S^2 - L$): we can interpret this as an oversampling of the $L$ links used to calculate this probability. This and other issues have been raised about the AUC~\cite{auc-misleading}, but we sidestep many of these complaints by using a deeper analysis of goodness of fit, including considering likelihood scores, the variance created by disparate sizes of link and non-link sets, and performance of the model on other network properties.  Finally, the AUC integrates over the performance of the model over all probability thresholds, describing a wider range of goodness of fit, and it is a more appropriate measure than error rate (or fraction of links predicted~\cite{PNM2011}) to emphasize the model's goodness of fit on the true links.

\section*{S3. Parallel tempering for optimization.}

\subsection*{Optimization techniques for finding the maximum likelihood of the PNM}

Recall that the likelihood for the PNM was defined as:
\begin{eqnarray}\label{eqn:pnmlikelihood}
\mathcal{L}(G | \theta) &=& \sum_{i=1}^S \sum_{j=1}^S  \ln\left\{  \begin{array}{l l}
\Pr(i \rightarrow j | \theta) & \text{ if }G(i,j) = 1 \\
1 - \Pr(i \rightarrow j | \theta) & \text{ otherwise }
\end{array} \right. \\
\end{eqnarray}
where
\begin{eqnarray}
\Pr(i \rightarrow j|\theta) &=& \alpha \exp\left( - \left|\frac{n_j - c_i}{r_i/2} \right|^e \right).
\end{eqnarray}
When we fit the model parameters from the data, we aim to maximize this objective function. However, for the PNM, there are many inherent symmetries in the parameter space and there are many  local optima. That is, for the PNM, the likelihood space is rugged: to search within this space for optima, there are many paths and relationships between the parameters that can lead to local optima and poorly fitted model parameters. Due to this ruggedness, it is nontrivial to find good maximum likelihood estimates of the parameters for the PNM given the data. \textit{Ad hoc}, metadata-driven heuristics have been successfully applied to more quickly find good estimates, but these methods do not scale well or provide guidance in the absence of sufficient metadata. Instead, we suggest parallel tempering to find reasonably good estimates using an efficient, easily parallelizable algorithm~\cite{paralleltemp-earl-deem}.

Parallel tempering, also known as replica exchange Markov Chain Monte Carlo (replica exchange MCMC), is an optimization technique from statistical physics. To our knowledge, it has never been applied in ecology, and it serves as a powerful optimization tool for working with probabilistic models. It is a meta-MCMC algorithm, in the sense that it combines several MCMC chains into a larger chain, while maintaining detailed balance~\cite{paralleltemp-earl-deem, newman-barkema-MCMC}. By mixing high temperature (fast-mixing) and low temperature (slow-mixing, locally hill-climbing) chains, parallel tempering allows us to more quickly survey the likelihood space and explore more diverse local optima (Figure 4, main text). This robustness is particularly useful in this setting, where food web datasets are often unevenly resolved data and where the models have many parameters.

 Parallel tempering employs $Q$ Markov chains in parallel, over a range of mixing temperatures $T_1, \ldots, T_Q$. For experiments described in the paper, we use $Q=35$. Each of these chains uses the Metropolis-Hastings algorithm internally for a fixed number of steps. Then, after a fixed number of steps (we use \texttt{NSTEPS} = 1000), a pair of chains at adjacent temperatures $T_k, T_{k+1}$ are randomly chosen. Following a Metropolis-Hastings update rule, $\Pr(switch) = \min\{1, \exp((\frac{1}{T_i} - \frac{1}{T_j})(\mathcal{L}_i - \mathcal{L}_j)) \}$, these chains are allowed to `switch' states~\cite{paralleltemp-earl-deem,newman-barkema-MCMC}. Specifically, this means that while all other chains resume searching the space from the same state that they ended at, the `switched' chains start searching from the other's state. The different-temperature chains explore the space at different rates, and by switching states, different parts of the space can be explored at different rates, for example, by discovering a new region with a high-temperature chain and then locally exploring with a low-temperature chain. 

There exist a number of methods to fine-tune the choice of the set of mixing temperatures, as well as the length and number of epochs, however, since these parameters might vary widely by application and create a high computational cost and technical barrier for most users. Reasonable parameters can be chosen to ensure reasonable mixing and performance, we suggest some heuristics and reasonable defaults for these choices in our code.

\section*{S4. Evidence for models and for type differences.}

\subsection*{S4.1 Testing for the inverse niche model and the cascade model in the parameters of the PNM}
The flexibility of the PNM allows it to recover various models and hypotheses about the structure of food webs. The PNM encodes these models and hypotheses through certain configurations of parameter settings. Here we present several tests of the cascade model and the inverse niche model as contained by the PNM. By making these niche model probabilistic, it handles potentially quasi-interval webs, as has been empirically successful previously~\cite{PNM2010,PNM2011,Zook2010}. We take these previous investigations, as well as our evaluations of goodness of fit and model performance, as evidence for a more general, probabilistic variant of the original niche model.

In this section and section S4.2, all results shown are calculated from optima found from each of 100 different parallel tempering runs, as used in the main text.

\paragraph{The cascade model}
The cascade model assumes that there exists an ordering on the one dimensional niche space (niche axis) such that every species consumes only species lower than themselves in that space. Then, the cascade model would be recovered by the PNM with the parameters $n_{i} \ge c_{i}$ for each species. That is, each species' niche position is higher than the center of their feeding range. It is worthwhile to note that the cascade model as presented by Ref.~\cite{cattin2004phylogenetic} is more strict than encoding the cascade model within the PNM. Firstly, the PNM version allows the cascade ordering to be probabilistic. In addition, even if the center of the feeding range was below the niche position, the form of the feeding distribution would still allow species to consume resources with higher niche positions with low probability.  

We test the cascade model both in a strict and approximate form. If the model recovered by the PNM strictly follows the cascade model, we would see $n_{i} \ge c_{i}$ for all $i$. We also take a softer test, and look for the the average value $n_{i} - c_{i}$. If, in general, $n_i \ge c_i$, then we expect this quantity to be positive. This characterize the the average difference of the niche positions of consumers and their most likely resources.

We find there is no evidence for the cascade model in the strict sense (Table S4). That is, it was never true that for every species, $n_i \ge c_i$.  We also did not find strong evidence of the approximate cascade model (Table S4). For each web, the average $(n_{i} - c_{i})$ was not statistically significantly different than zero (and specifically, we can not say definitively if the true value is positive or negative). This means that the PNM recovers little evidence for the cascade model.

\begin{table*}
\centering
\begin{ruledtabular}
\begin{tabular}{l|ccc}
&  $\mathbf{G_{F}}$ &  $\mathbf{G_{FP}}$ &  $\mathbf{G_{FPC}}$\\\hline
Cascade model, strict 			& False (0/100)		& False (0/100)		& False (0/100) \\
Cascade model, weak	&&&\\
$\ \ \ \ \text{mean}(n_{i} - c_{i})\ (\text{std\ err}), \forall i$	& 0.025 (0.080)		& 0.020 (0.061) & 0.026 (0066) \\ \hline 
Inverse niche model, strict		& ---	& False (0/100)		& False (0/100)	\\ 
Inverse niche model, weak &&&\\
$\ \ \ \ \text{mean}(n_{i} - c_{i})\ (\text{std\ err}),\ i \in V_{F} $ &	---		& 0.037 (0.068) 	& 0.048 (0.068)		\\
$\ \ \ \ \text{mean}(n_{i} - c_{i})\ (\text{std\ err}),\ i \in V_{P} $ &	---		& -0.009 (0.098)	& -0.011 (0.107)	\\
$\ \ \ \ $Correlation $\rho(n_{i},r_{i}),\ i \in V_{P}$ 		    & ---	 	& -0.098 	& -0.006  	\\
\end{tabular}
\end{ruledtabular}
\captionsetup{labelformat=empty,justification=centering}
\caption{{\bf Table S4} Evidence for the cascade model and the inverse niche model}
\label{tab:cascadeInv}
\end{table*}

\paragraph{The inverse niche model}
The inverse niche model is defined for free-living predation $E_{F\rightarrow F}$ and parasitism on free-living species $E_{P\rightarrow F}$~\cite{inverse-niche-model}. For free-living species, the model follows the cascade model, where species consume on species lower on the niche axis than themselves. However, parasites follow an inverse cascade, ordered such that they only consume species higher than themselves on the niche axis, $n_{i} \le c_{i}$. In addition, the inverse niche model requires that the parasite's feeding range $r_{i}$ is negatively correlated with its niche position $n_{i}$. 

We test for the inverse niche model both in a strict and approximate form. If the model strictly followed the inverse niche model, we would see, for all free-living species $i$, $n_{i} \ge c_{i}$, and for all parasites $k$, $n_{k} \le c_{k}$ and $n_{k}$ negatively correlated with $r_{k}$. Again, this is still a forgiving test: these are still probabilistic orderings, and species have wide enough feeding distributions that they can consume species higher (resp., lower) than themselves with nonzero probability. 

We find no evidence for the inverse niche model in the strict sense (Table S4). That is, it was never true that for every free-living species, $n_i \ge c_i$, and vice-versa for parasites. We also find no evidence from the approximate test: for both free-living species and parasites, the niche difference $n_{i} - c_{i}$ is not statistically significantly different from zero, and we can not definitively say that either is positive (for free-living species) or negative (for parasites). In addition, there is effectively zero correlation between the niche position and the feeding range width for parasites (Table S4). This suggests there is no direct evidence for the inverse niche model from these results.

\subsection*{S4.2 Are parasites and free-living species different?}

 We compare the empirical distributions of the parameters for free-living species and parasites. We test this on the models fit to both webs with parasites, $G_{FP}$ and $G_{FPC}$; this also allows us to test whether or not observing concomitant links changes these results. 

 We use the Kolmogorov-Smirnov (KS) statistic to test the null hypothesis that the parameters for free-living species were drawn from the same distribution as the parameters for parasites. The KS test is a nonparametric statistical tool to compare two distributions. Here, we are interested in comparing two empirical distribution functions, that of the parameters from free-living species against the parameters from parasites. The statistic of interest is the p-value: if the p-value is small, we can reject the null hypothesis that the two distributions came from the same underlying distribution. Conversely, if the p-value is large, we fail to reject the null hypothesis that the two distributions came from the same underlying distribution. Here we account for having samples of different sizes---there are more free-living species than parasite species---in the calculation of the p-value. 

\begin{table*}
\centering
\begin{ruledtabular}
\begin{tabular}{l|lccl}
				& {\bf Data} 			& {\bf KS statistic} & {\bf p-value} 	& {\bf Reject null?} \\\hline 
Niche position $n_{i}$ & $G_{FP}$ 		& 0.169	& $\ll 0.001$	& Yes \\
Niche position $n_{i}$ & $G_{FPC}$	 	& 0.056	& $\ll 0.001$	& Yes \\\hline
Feeding position $c_{i}$ & $G_{FP}$ 	& 0.115	& $\ll 0.001$	& Yes \\
Feeding position $c_{i}$ & $G_{FPC}$ 	& 0.122	& $\ll 0.001$	& Yes \\\hline
Feeding range width $r_{i}$ & $G_{FP}$ 	& 0.046	& $\ll 0.001$	& Yes \\
Feeding range width $r_{i}$ & $G_{FPC}$ & 0.080	& $\ll 0.001$	& Yes \\\hline
Cascade difference $n_{i} - c_{i}$ & $G_{FP}$ 		& 0.428 &   $\ll 0.001$ & Yes \\
Cascade difference $n_{i} - c_{i}$ & $G_{FPC}$ 	& 0.207 & 0.198	& No \\\hline
Niche-width relation $n_{i}/r_{i}$ & $G_{FP}$ 		& 0.151 & 0.568 & No \\
Niche-width relation $n_{i}/r_{i}$ & $G_{FPC}$ 		& 0.128 & 0.764 & No \\
\end{tabular}
\end{ruledtabular}
\captionsetup{labelformat=empty,justification=centering}
\caption{{\bf Table S5} Comparisons of distributions of parameters of free-living vs.\ parasite taxa}
\label{tab:KStests}
\end{table*}

We find that all species-specific parameters, $n_{i},\ c_{i},\ r_{i}$ appear to come from different distributions for parasites vs.\ free-living species (Table S5). That is, both free-living species and parasites are typically distributed differently in the niche space (here, defined to be the interval $[0,1]$). This is true whether or not concomitant links are included.

Recall that the difference between the niche position and the feeding center, $n_{i} - c_{i}$, served to approximately represent the cascade and inverse niche model. When concomitant links were not observed, these distributions are statistically significantly different (Table S5). However, when concomitant links are observed ($G_{FPC}$), these distributions are not different. Regardless of concomitant links, the scaling relationship between the niche position and the feeding width, $n_{i}/r_{i}$, is not significantly different for free-living and parasite species.

\section*{S5. Motifs and the PNM.}

\subsection*{Intraguild trophic interactions among parasites help the PNM represent concomitant links}

While structural motifs and niches represent food web structure at fundamentally different levels, the distribution of motifs present in a food web affects how the PNM will fit that web. Based on structural motifs, Cirtwill and Stouffer~\cite{cirtwill2015concomitant} found that parasites' roles are more similar to intermediate free-living consumers with a number of both predators and prey. However, when concomitant links were included, parasites roles were different from all other types of species. Furthermore, they found that concomitant links are the most structurally diverse type of interaction. 

In the context of the PNM, the local structure of motifs can give the model information about how taxa are related to each other in the niche space. For example, one species consuming another does not provide information about their locations in the niche space. However, if the consumption was reciprocal, then we would know more: for each taxa, their niche position would be near the center of the other's feeding range, so both the niche positions and feeding ranges would be paired. Expanding to more complex relationships, the diversity of a species' diet can be reflected in the width of their feeding distribution. However, species being consumed together means they will then be closer together in the niche space, to increase the likeilhood they are consumed at the same time. 

Considering the setting of parasites in food webs, the distribution of motifs will be related to the types of interactions observed when parasites or concomitant links are introduced~\cite{cirtwill2015concomitant}. Both concomitant predation and trophic intraguild predation occur in settings where a species will consume multiple resources, which themselves may feed on each other. When this happens, these relationships are represented by triangles in a web. Concomitant predation induces a greater distribution of these triangles, and thus a different motif distribution for these webs. 

\begin{figure*}[thbp]
\centering
\includegraphics[width=4in]{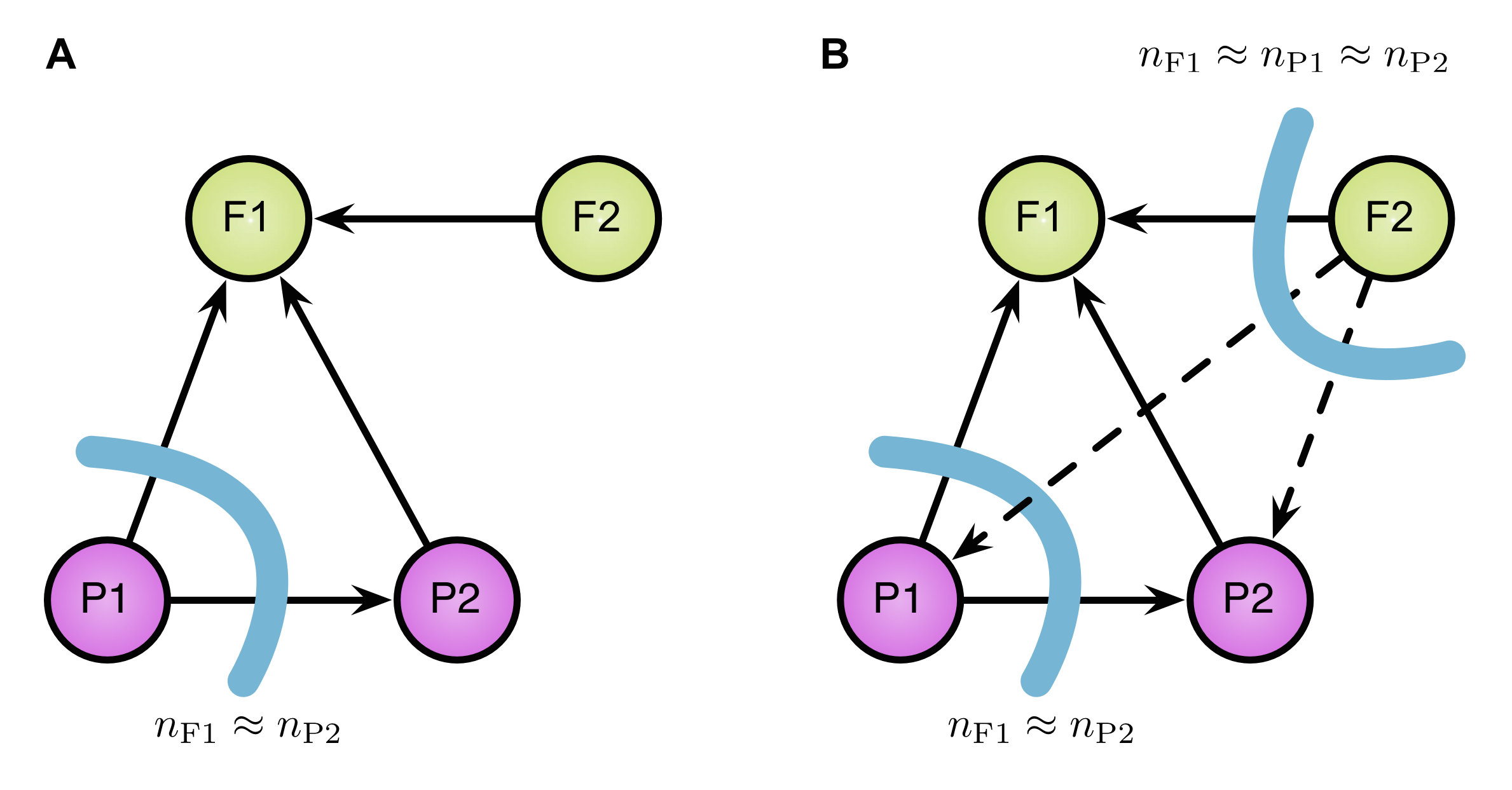}
\captionsetup{labelformat=empty,justification=centering} 
\caption{{\bf Figure S3.} Intraguild trophic interactions among parasites in the PNM supports the model's ability to predict concomitant links. A) Under the PNM, the likelihood is maximized when resources with the same consumer have more similar niche positions. B) When concomitant links (dashed arrows) are introduced, there is even more evidence (more links) that the resources' niche positions should be closer. It increases the likelihood of the model when the parasites and their host have similar niche positions. 
}\label{fig:cirtwilltriangles}
\end{figure*}

We illustrate an example of how trophic interactions among parasites can help predict concomitant links in Figure S3
with two free-living species and two parasites. We consider fitting the PNM to such a web in which a parasite consumes another parasite within-host. Since the consumed parasite and the host are consumed together, their niche positions will be pushed together. If some other free-living species consumes the host, then at least one of the parasites infecting the host will be more likely to be consumed, assigning higher probability to concomitant predation links. Conversely, if concomitant predation links are also included in the web, then the niche positions of parasites and their host will be closer together in the niche space. Then the model will predict that those parasites feed in that region of the niche space where their host is with high probability, and also compete with the other parasites sharing that host.

We observe the effects of this in the link prediction task, described in the main text. As we see in Table 3 of the main text, it is easier to recover links in $E_{PP}$ when we have also observed concomitant links, i.e., having observed a noisy version of $G_{FPC}$ rather than $G_{FP}$. That is, it is easier to predict trophic interactions among parasites when we have observed concomitant predation links. Concomitant predation links provide evidence to the model of parasites that would share hosts and thus compete, allowing even unobserved intraguild predation links to be predicted effectively. This suggests that concomitant links naturally encode a niche intimacy between parasites and their hosts in the PNM.

\end{bibunit}


\begin{thebibliography}{34}%
\makeatletter
\providecommand \@ifxundefined [1]{%
 \@ifx{#1\undefined}
}%
\providecommand \@ifnum [1]{%
 \ifnum #1\expandafter \@firstoftwo
 \else \expandafter \@secondoftwo
 \fi
}%
\providecommand \@ifx [1]{%
 \ifx #1\expandafter \@firstoftwo
 \else \expandafter \@secondoftwo
 \fi
}%
\providecommand \natexlab [1]{#1}%
\providecommand \enquote  [1]{``#1''}%
\providecommand \bibnamefont  [1]{#1}%
\providecommand \bibfnamefont [1]{#1}%
\providecommand \citenamefont [1]{#1}%
\providecommand \href@noop [0]{\@secondoftwo}%
\providecommand \href [0]{\begingroup \@sanitize@url \@href}%
\providecommand \@href[1]{\@@startlink{#1}\@@href}%
\providecommand \@@href[1]{\endgroup#1\@@endlink}%
\providecommand \@sanitize@url [0]{\catcode `\\12\catcode `\$12\catcode
  `\&12\catcode `\#12\catcode `\^12\catcode `\_12\catcode `\%12\relax}%
\providecommand \@@startlink[1]{}%
\providecommand \@@endlink[0]{}%
\providecommand \url  [0]{\begingroup\@sanitize@url \@url }%
\providecommand \@url [1]{\endgroup\@href {#1}{\urlprefix }}%
\providecommand \urlprefix  [0]{URL }%
\providecommand \Eprint [0]{\href }%
\providecommand \doibase [0]{http://dx.doi.org/}%
\providecommand \selectlanguage [0]{\@gobble}%
\providecommand \bibinfo  [0]{\@secondoftwo}%
\providecommand \bibfield  [0]{\@secondoftwo}%
\providecommand \translation [1]{[#1]}%
\providecommand \BibitemOpen [0]{}%
\providecommand \bibitemStop [0]{}%
\providecommand \bibitemNoStop [0]{.\EOS\space}%
\providecommand \EOS [0]{\spacefactor3000\relax}%
\providecommand \BibitemShut  [1]{\csname bibitem#1\endcsname}%
\let\auto@bib@innerbib\@empty
\bibitem [{\citenamefont {Hudson}\ \emph {et~al.}(2006)\citenamefont {Hudson},
  \citenamefont {Dobson},\ and\ \citenamefont {Lafferty}}]{hudson2006healthy}%
  \BibitemOpen
  \bibfield  {author} {\bibinfo {author} {\bibfnamefont {P.~J.}\ \bibnamefont
  {Hudson}}, \bibinfo {author} {\bibfnamefont {A.~P.}\ \bibnamefont {Dobson}},
  \ and\ \bibinfo {author} {\bibfnamefont {K.~D.}\ \bibnamefont {Lafferty}},\
  }\href@noop {} {\bibfield  {journal} {\bibinfo  {journal} {Trends in Ecology
  \& Evolution}\ }\textbf {\bibinfo {volume} {21}},\ \bibinfo {pages} {381}
  (\bibinfo {year} {2006})}\BibitemShut {NoStop}%
\bibitem [{\citenamefont {Huxham}\ \emph {et~al.}(1996)\citenamefont {Huxham},
  \citenamefont {Beaney},\ and\ \citenamefont
  {Raffaelli}}]{huxham1996triangles}%
  \BibitemOpen
  \bibfield  {author} {\bibinfo {author} {\bibfnamefont {M.}~\bibnamefont
  {Huxham}}, \bibinfo {author} {\bibfnamefont {S.}~\bibnamefont {Beaney}}, \
  and\ \bibinfo {author} {\bibfnamefont {D.}~\bibnamefont {Raffaelli}},\
  }\href@noop {} {\bibfield  {journal} {\bibinfo  {journal} {Oikos}\ ,\
  \bibinfo {pages} {284}} (\bibinfo {year} {1996})}\BibitemShut {NoStop}%
\bibitem [{\citenamefont {Marcogliese}\ and\ \citenamefont
  {Cone}(1997)}]{Marcogliese-parasiteplea}%
  \BibitemOpen
  \bibfield  {author} {\bibinfo {author} {\bibfnamefont {D.~J.}\ \bibnamefont
  {Marcogliese}}\ and\ \bibinfo {author} {\bibfnamefont {D.~K.}\ \bibnamefont
  {Cone}},\ }\href {\doibase 10.1016/S0169-5347(97)01080-X} {\bibfield
  {journal} {\bibinfo  {journal} {Trends in Ecology and Evolution}\ }\textbf
  {\bibinfo {volume} {12}},\ \bibinfo {pages} {320 } (\bibinfo {year}
  {1997})}\BibitemShut {NoStop}%
\bibitem [{\citenamefont {Lafferty}\ \emph {et~al.}(2008)\citenamefont
  {Lafferty}, \citenamefont {Allesina}, \citenamefont {Arim}, \citenamefont
  {Briggs}, \citenamefont {Leo}, \citenamefont {Dobson}, \citenamefont {Dunne},
  \citenamefont {Johnson}, \citenamefont {Kuris}, \citenamefont {Marcogliese},
  \citenamefont {Martinez}, \citenamefont {Memmott}, \citenamefont {Marquet},
  \citenamefont {McLaughlin}, \citenamefont {Mordecai}, \citenamefont
  {Pascual}, \citenamefont {Poulin},\ and\ \citenamefont
  {Thieltges}}]{para-missing-links}%
  \BibitemOpen
  \bibfield  {author} {\bibinfo {author} {\bibfnamefont {K.~D.}\ \bibnamefont
  {Lafferty}}, \bibinfo {author} {\bibfnamefont {S.}~\bibnamefont {Allesina}},
  \bibinfo {author} {\bibfnamefont {M.}~\bibnamefont {Arim}}, \bibinfo {author}
  {\bibfnamefont {C.~J.}\ \bibnamefont {Briggs}}, \bibinfo {author}
  {\bibfnamefont {G.~D.}\ \bibnamefont {Leo}}, \bibinfo {author} {\bibfnamefont
  {A.~P.}\ \bibnamefont {Dobson}}, \bibinfo {author} {\bibfnamefont {J.~A.}\
  \bibnamefont {Dunne}}, \bibinfo {author} {\bibfnamefont {P.~T.~J.}\
  \bibnamefont {Johnson}}, \bibinfo {author} {\bibfnamefont {A.~M.}\
  \bibnamefont {Kuris}}, \bibinfo {author} {\bibfnamefont {D.~J.}\ \bibnamefont
  {Marcogliese}}, \bibinfo {author} {\bibfnamefont {N.~D.}\ \bibnamefont
  {Martinez}}, \bibinfo {author} {\bibfnamefont {J.}~\bibnamefont {Memmott}},
  \bibinfo {author} {\bibfnamefont {P.~A.}\ \bibnamefont {Marquet}}, \bibinfo
  {author} {\bibfnamefont {J.~P.}\ \bibnamefont {McLaughlin}}, \bibinfo
  {author} {\bibfnamefont {E.~A.}\ \bibnamefont {Mordecai}}, \bibinfo {author}
  {\bibfnamefont {M.}~\bibnamefont {Pascual}}, \bibinfo {author} {\bibfnamefont
  {R.}~\bibnamefont {Poulin}}, \ and\ \bibinfo {author} {\bibfnamefont {D.~W.}\
  \bibnamefont {Thieltges}},\ }\href@noop {} {\bibfield  {journal} {\bibinfo
  {journal} {Ecology Letters}\ }\textbf {\bibinfo {volume} {11}},\ \bibinfo
  {pages} {533} (\bibinfo {year} {2008})}\BibitemShut {NoStop}%
\bibitem [{\citenamefont {Johnson}\ \emph {et~al.}(2010)\citenamefont
  {Johnson}, \citenamefont {Dobson}, \citenamefont {Lafferty}, \citenamefont
  {Marcogliese}, \citenamefont {Memmott}, \citenamefont {Orlofske},
  \citenamefont {Poulin},\ and\ \citenamefont
  {Thieltges}}]{johnson-parasites-prey}%
  \BibitemOpen
  \bibfield  {author} {\bibinfo {author} {\bibfnamefont {P.~T.~J.}\
  \bibnamefont {Johnson}}, \bibinfo {author} {\bibfnamefont {A.}~\bibnamefont
  {Dobson}}, \bibinfo {author} {\bibfnamefont {K.~D.}\ \bibnamefont
  {Lafferty}}, \bibinfo {author} {\bibfnamefont {D.~J.}\ \bibnamefont
  {Marcogliese}}, \bibinfo {author} {\bibfnamefont {J.}~\bibnamefont
  {Memmott}}, \bibinfo {author} {\bibfnamefont {S.~A.}\ \bibnamefont
  {Orlofske}}, \bibinfo {author} {\bibfnamefont {R.}~\bibnamefont {Poulin}}, \
  and\ \bibinfo {author} {\bibfnamefont {D.~W.}\ \bibnamefont {Thieltges}},\
  }\href@noop {} {\bibfield  {journal} {\bibinfo  {journal} {Trends in Ecology
  and Evolution}\ }\textbf {\bibinfo {volume} {25}},\ \bibinfo {pages} {362}
  (\bibinfo {year} {2010})}\BibitemShut {NoStop}%
\bibitem [{\citenamefont {Fontaine}\ \emph {et~al.}(2011)\citenamefont
  {Fontaine}, \citenamefont {Guimar{\~a}es}, \citenamefont {K{\'e}fi},
  \citenamefont {Loeuille}, \citenamefont {Memmott}, \citenamefont {Van
  Der~Putten}, \citenamefont {Van~Veen},\ and\ \citenamefont
  {Th{\'e}bault}}]{fontaine2011ecological}%
  \BibitemOpen
  \bibfield  {author} {\bibinfo {author} {\bibfnamefont {C.}~\bibnamefont
  {Fontaine}}, \bibinfo {author} {\bibfnamefont {P.~R.}\ \bibnamefont
  {Guimar{\~a}es}}, \bibinfo {author} {\bibfnamefont {S.}~\bibnamefont
  {K{\'e}fi}}, \bibinfo {author} {\bibfnamefont {N.}~\bibnamefont {Loeuille}},
  \bibinfo {author} {\bibfnamefont {J.}~\bibnamefont {Memmott}}, \bibinfo
  {author} {\bibfnamefont {W.~H.}\ \bibnamefont {Van Der~Putten}}, \bibinfo
  {author} {\bibfnamefont {F.~J.}\ \bibnamefont {Van~Veen}}, \ and\ \bibinfo
  {author} {\bibfnamefont {E.}~\bibnamefont {Th{\'e}bault}},\ }\href@noop {}
  {\bibfield  {journal} {\bibinfo  {journal} {Ecology Letters}\ }\textbf
  {\bibinfo {volume} {14}},\ \bibinfo {pages} {1170} (\bibinfo {year}
  {2011})}\BibitemShut {NoStop}%
\bibitem [{\citenamefont {K{\'e}fi}\ \emph {et~al.}(2012)\citenamefont
  {K{\'e}fi}, \citenamefont {Berlow}, \citenamefont {Wieters}, \citenamefont
  {Navarrete}, \citenamefont {Petchey}, \citenamefont {Wood}, \citenamefont
  {Boit}, \citenamefont {Joppa}, \citenamefont {Lafferty}, \citenamefont
  {Williams} \emph {et~al.}}]{kefi2012merging}%
  \BibitemOpen
  \bibfield  {author} {\bibinfo {author} {\bibfnamefont {S.}~\bibnamefont
  {K{\'e}fi}}, \bibinfo {author} {\bibfnamefont {E.~L.}\ \bibnamefont
  {Berlow}}, \bibinfo {author} {\bibfnamefont {E.~A.}\ \bibnamefont {Wieters}},
  \bibinfo {author} {\bibfnamefont {S.~A.}\ \bibnamefont {Navarrete}}, \bibinfo
  {author} {\bibfnamefont {O.~L.}\ \bibnamefont {Petchey}}, \bibinfo {author}
  {\bibfnamefont {S.~A.}\ \bibnamefont {Wood}}, \bibinfo {author}
  {\bibfnamefont {A.}~\bibnamefont {Boit}}, \bibinfo {author} {\bibfnamefont
  {L.~N.}\ \bibnamefont {Joppa}}, \bibinfo {author} {\bibfnamefont {K.~D.}\
  \bibnamefont {Lafferty}}, \bibinfo {author} {\bibfnamefont {R.~J.}\
  \bibnamefont {Williams}},  \emph {et~al.},\ }\href@noop {} {\bibfield
  {journal} {\bibinfo  {journal} {Ecology Letters}\ }\textbf {\bibinfo {volume}
  {15}},\ \bibinfo {pages} {291} (\bibinfo {year} {2012})}\BibitemShut
  {NoStop}%
\bibitem [{\citenamefont {Britton}(2013)}]{britton-2013}%
  \BibitemOpen
  \bibfield  {author} {\bibinfo {author} {\bibfnamefont {J.~R.}\ \bibnamefont
  {Britton}},\ }\href@noop {} {\bibfield  {journal} {\bibinfo  {journal}
  {Trends in Ecology and Evolution}\ }\textbf {\bibinfo {volume} {28}},\
  \bibinfo {pages} {93} (\bibinfo {year} {2013})}\BibitemShut {NoStop}%
\bibitem [{\citenamefont {Thompson}\ \emph {et~al.}(2005)\citenamefont
  {Thompson}, \citenamefont {Mouritsen},\ and\ \citenamefont
  {Poulin}}]{thompson2005importance}%
  \BibitemOpen
  \bibfield  {author} {\bibinfo {author} {\bibfnamefont {R.~M.}\ \bibnamefont
  {Thompson}}, \bibinfo {author} {\bibfnamefont {K.~N.}\ \bibnamefont
  {Mouritsen}}, \ and\ \bibinfo {author} {\bibfnamefont {R.}~\bibnamefont
  {Poulin}},\ }\href@noop {} {\bibfield  {journal} {\bibinfo  {journal}
  {Journal of Animal Ecology}\ }\textbf {\bibinfo {volume} {74}},\ \bibinfo
  {pages} {77} (\bibinfo {year} {2005})}\BibitemShut {NoStop}%
\bibitem [{\citenamefont {Hernandez}\ and\ \citenamefont
  {Sukhdeo}(2008)}]{hernandez2008parasites}%
  \BibitemOpen
  \bibfield  {author} {\bibinfo {author} {\bibfnamefont {A.~D.}\ \bibnamefont
  {Hernandez}}\ and\ \bibinfo {author} {\bibfnamefont {M.~V.}\ \bibnamefont
  {Sukhdeo}},\ }\href@noop {} {\bibfield  {journal} {\bibinfo  {journal}
  {Oecologia}\ }\textbf {\bibinfo {volume} {156}},\ \bibinfo {pages} {613}
  (\bibinfo {year} {2008})}\BibitemShut {NoStop}%
\bibitem [{\citenamefont {Kuang}\ and\ \citenamefont
  {Zhang}(2011)}]{kuangzhang2011}%
  \BibitemOpen
  \bibfield  {author} {\bibinfo {author} {\bibfnamefont {W.}~\bibnamefont
  {Kuang}}\ and\ \bibinfo {author} {\bibfnamefont {W.}~\bibnamefont {Zhang}},\
  }\href@noop {} {\bibfield  {journal} {\bibinfo  {journal} {Network Biology}\
  }\textbf {\bibinfo {volume} {1}},\ \bibinfo {pages} {171} (\bibinfo {year}
  {2011})}\BibitemShut {NoStop}%
\bibitem [{\citenamefont {Thompson}\ \emph {et~al.}(2012)\citenamefont
  {Thompson}, \citenamefont {Brose}, \citenamefont {Dunne}, \citenamefont
  {Hall}, \citenamefont {Hladyz}, \citenamefont {Kitching}, \citenamefont
  {Martinez}, \citenamefont {Rantala}, \citenamefont {Romanuk}, \citenamefont
  {Stouffer},\ and\ \citenamefont {Tylianakis}}]{fw-structurefn-biodiversity}%
  \BibitemOpen
  \bibfield  {author} {\bibinfo {author} {\bibfnamefont {R.~M.}\ \bibnamefont
  {Thompson}}, \bibinfo {author} {\bibfnamefont {U.}~\bibnamefont {Brose}},
  \bibinfo {author} {\bibfnamefont {J.~A.}\ \bibnamefont {Dunne}}, \bibinfo
  {author} {\bibfnamefont {R.~O.~J.}\ \bibnamefont {Hall}}, \bibinfo {author}
  {\bibfnamefont {S.}~\bibnamefont {Hladyz}}, \bibinfo {author} {\bibfnamefont
  {R.~L.}\ \bibnamefont {Kitching}}, \bibinfo {author} {\bibfnamefont {N.~D.}\
  \bibnamefont {Martinez}}, \bibinfo {author} {\bibfnamefont {H.}~\bibnamefont
  {Rantala}}, \bibinfo {author} {\bibfnamefont {T.~N.}\ \bibnamefont
  {Romanuk}}, \bibinfo {author} {\bibfnamefont {D.~B.}\ \bibnamefont
  {Stouffer}}, \ and\ \bibinfo {author} {\bibfnamefont {J.~M.}\ \bibnamefont
  {Tylianakis}},\ }\href@noop {} {\bibfield  {journal} {\bibinfo  {journal}
  {Trends in Ecology and Evolution}\ } (\bibinfo {year} {2012})}\BibitemShut
  {NoStop}%
\bibitem [{\citenamefont {Warren}\ \emph {et~al.}(2010)\citenamefont {Warren},
  \citenamefont {Pascual}, \citenamefont {Lafferty},\ and\ \citenamefont
  {Kuris}}]{inverse-niche-model}%
  \BibitemOpen
  \bibfield  {author} {\bibinfo {author} {\bibfnamefont {C.~P.}\ \bibnamefont
  {Warren}}, \bibinfo {author} {\bibfnamefont {M.}~\bibnamefont {Pascual}},
  \bibinfo {author} {\bibfnamefont {K.~D.}\ \bibnamefont {Lafferty}}, \ and\
  \bibinfo {author} {\bibfnamefont {A.~M.}\ \bibnamefont {Kuris}},\ }\href@noop
  {} {\bibfield  {journal} {\bibinfo  {journal} {Theoretical Ecology}\ }\textbf
  {\bibinfo {volume} {3}},\ \bibinfo {pages} {285} (\bibinfo {year}
  {2010})}\BibitemShut {NoStop}%
\bibitem [{\citenamefont {Dunne}\ \emph {et~al.}(2013)\citenamefont {Dunne},
  \citenamefont {Lafferty}, \citenamefont {Dobson}, \citenamefont {Hechinger},
  \citenamefont {Kuris}, \citenamefont {Martinez}, \citenamefont {McLaughlin},
  \citenamefont {Mouritsen}, \citenamefont {Poulin}, \citenamefont {Reise}
  \emph {et~al.}}]{dunne-2013-parafw}%
  \BibitemOpen
  \bibfield  {author} {\bibinfo {author} {\bibfnamefont {J.~A.}\ \bibnamefont
  {Dunne}}, \bibinfo {author} {\bibfnamefont {K.~D.}\ \bibnamefont {Lafferty}},
  \bibinfo {author} {\bibfnamefont {A.~P.}\ \bibnamefont {Dobson}}, \bibinfo
  {author} {\bibfnamefont {R.~F.}\ \bibnamefont {Hechinger}}, \bibinfo {author}
  {\bibfnamefont {A.~M.}\ \bibnamefont {Kuris}}, \bibinfo {author}
  {\bibfnamefont {N.~D.}\ \bibnamefont {Martinez}}, \bibinfo {author}
  {\bibfnamefont {J.~P.}\ \bibnamefont {McLaughlin}}, \bibinfo {author}
  {\bibfnamefont {K.~N.}\ \bibnamefont {Mouritsen}}, \bibinfo {author}
  {\bibfnamefont {R.}~\bibnamefont {Poulin}}, \bibinfo {author} {\bibfnamefont
  {K.}~\bibnamefont {Reise}},  \emph {et~al.},\ }\href@noop {} {\bibfield
  {journal} {\bibinfo  {journal} {PLOS biology}\ }\textbf {\bibinfo {volume}
  {11}},\ \bibinfo {pages} {e1001579} (\bibinfo {year} {2013})}\BibitemShut
  {NoStop}%
\bibitem [{\citenamefont {Martinez}\ and\ \citenamefont
  {Dunne}(1998)}]{martinez1998time}%
  \BibitemOpen
  \bibfield  {author} {\bibinfo {author} {\bibfnamefont {N.}~\bibnamefont
  {Martinez}}\ and\ \bibinfo {author} {\bibfnamefont {J.~A.}\ \bibnamefont
  {Dunne}},\ }in\ \href@noop {} {\emph {\bibinfo {booktitle} {Ecological Scale:
  Theory and Applications}}},\ \bibinfo {editor} {edited by\ \bibinfo {editor}
  {\bibfnamefont {D.}~\bibnamefont {Peterson}}\ and\ \bibinfo {editor}
  {\bibfnamefont {V.}~\bibnamefont {Parker}}}\ (\bibinfo  {publisher} {Columbia
  University Press},\ \bibinfo {year} {1998})\ pp.\ \bibinfo {pages}
  {207--226}\BibitemShut {NoStop}%
\bibitem [{\citenamefont {Stouffer}\ \emph {et~al.}(2007)\citenamefont
  {Stouffer}, \citenamefont {Camacho}, \citenamefont {Jiang},\ and\
  \citenamefont {Amaral}}]{stouffer2007evidence}%
  \BibitemOpen
  \bibfield  {author} {\bibinfo {author} {\bibfnamefont {D.~B.}\ \bibnamefont
  {Stouffer}}, \bibinfo {author} {\bibfnamefont {J.}~\bibnamefont {Camacho}},
  \bibinfo {author} {\bibfnamefont {W.}~\bibnamefont {Jiang}}, \ and\ \bibinfo
  {author} {\bibfnamefont {L.~A.~N.}\ \bibnamefont {Amaral}},\ }\href@noop {}
  {\bibfield  {journal} {\bibinfo  {journal} {Proceedings of the Royal Society
  B: Biological Sciences}\ }\textbf {\bibinfo {volume} {274}},\ \bibinfo
  {pages} {1931} (\bibinfo {year} {2007})}\BibitemShut {NoStop}%
\bibitem [{\citenamefont {Williams}\ and\ \citenamefont
  {Purves}(2011)}]{PNM2011}%
  \BibitemOpen
  \bibfield  {author} {\bibinfo {author} {\bibfnamefont {R.~J.}\ \bibnamefont
  {Williams}}\ and\ \bibinfo {author} {\bibfnamefont {D.~W.}\ \bibnamefont
  {Purves}},\ }\href@noop {} {\bibfield  {journal} {\bibinfo  {journal}
  {Ecology}\ }\textbf {\bibinfo {volume} {92}},\ \bibinfo {pages} {1849}
  (\bibinfo {year} {2011})}\BibitemShut {NoStop}%
\bibitem [{\citenamefont {Cirtwill}\ and\ \citenamefont
  {Stouffer}(2015)}]{cirtwill2015concomitant}%
  \BibitemOpen
  \bibfield  {author} {\bibinfo {author} {\bibfnamefont {A.~R.}\ \bibnamefont
  {Cirtwill}}\ and\ \bibinfo {author} {\bibfnamefont {D.~B.}\ \bibnamefont
  {Stouffer}},\ }\href {\doibase 10.1111/1365-2656.12323} {\bibfield  {journal}
  {\bibinfo  {journal} {Journal of Animal Ecology}\ }\textbf {\bibinfo {volume}
  {84}},\ \bibinfo {pages} {734} (\bibinfo {year} {2015})}\BibitemShut
  {NoStop}%
\bibitem [{\citenamefont {Williams}\ and\ \citenamefont
  {Martinez}(2000)}]{williams-martinez-niche-2000}%
  \BibitemOpen
  \bibfield  {author} {\bibinfo {author} {\bibfnamefont {R.~J.}\ \bibnamefont
  {Williams}}\ and\ \bibinfo {author} {\bibfnamefont {N.~D.}\ \bibnamefont
  {Martinez}},\ }\href@noop {} {\bibfield  {journal} {\bibinfo  {journal}
  {Nature}\ }\textbf {\bibinfo {volume} {404}},\ \bibinfo {pages} {180}
  (\bibinfo {year} {2000})}\BibitemShut {NoStop}%
\bibitem [{\citenamefont {Zander}\ \emph {et~al.}(2007)\citenamefont {Zander},
  \citenamefont {Josten}, \citenamefont {Detloff}, \citenamefont {Poulin},
  \citenamefont {McLaughlin},\ and\ \citenamefont
  {Thieltges}}]{flensburg-fjord-data}%
  \BibitemOpen
  \bibfield  {author} {\bibinfo {author} {\bibfnamefont {C.~D.}\ \bibnamefont
  {Zander}}, \bibinfo {author} {\bibfnamefont {N.}~\bibnamefont {Josten}},
  \bibinfo {author} {\bibfnamefont {K.~C.}\ \bibnamefont {Detloff}}, \bibinfo
  {author} {\bibfnamefont {R.}~\bibnamefont {Poulin}}, \bibinfo {author}
  {\bibfnamefont {J.~P.}\ \bibnamefont {McLaughlin}}, \ and\ \bibinfo {author}
  {\bibfnamefont {D.~W.}\ \bibnamefont {Thieltges}},\ }\href@noop {} {\bibfield
   {journal} {\bibinfo  {journal} {Ecology}\ }\textbf {\bibinfo {volume} {92}}
  (\bibinfo {year} {2007})}\BibitemShut {NoStop}%
\bibitem [{\citenamefont {Williams}\ \emph {et~al.}(2010)\citenamefont
  {Williams}, \citenamefont {Anandanadesan},\ and\ \citenamefont
  {Purves}}]{PNM2010}%
  \BibitemOpen
  \bibfield  {author} {\bibinfo {author} {\bibfnamefont {R.~J.}\ \bibnamefont
  {Williams}}, \bibinfo {author} {\bibfnamefont {A.}~\bibnamefont
  {Anandanadesan}}, \ and\ \bibinfo {author} {\bibfnamefont {D.}~\bibnamefont
  {Purves}},\ }\href {\doibase 10.1371/journal.pone.0012092} {\bibfield
  {journal} {\bibinfo  {journal} {PLOS ONE}\ }\textbf {\bibinfo {volume} {5}},\
  \bibinfo {pages} {e12092} (\bibinfo {year} {2010})}\BibitemShut {NoStop}%
\bibitem [{\citenamefont {Cattin}\ \emph {et~al.}(2004)\citenamefont {Cattin},
  \citenamefont {Bersier}, \citenamefont {Bana{\v{s}}ek-Richter}, \citenamefont
  {Baltensperger},\ and\ \citenamefont {Gabriel}}]{cattin2004phylogenetic}%
  \BibitemOpen
  \bibfield  {author} {\bibinfo {author} {\bibfnamefont {M.-F.}\ \bibnamefont
  {Cattin}}, \bibinfo {author} {\bibfnamefont {L.-F.}\ \bibnamefont {Bersier}},
  \bibinfo {author} {\bibfnamefont {C.}~\bibnamefont {Bana{\v{s}}ek-Richter}},
  \bibinfo {author} {\bibfnamefont {R.}~\bibnamefont {Baltensperger}}, \ and\
  \bibinfo {author} {\bibfnamefont {J.-P.}\ \bibnamefont {Gabriel}},\
  }\href@noop {} {\bibfield  {journal} {\bibinfo  {journal} {Nature}\ }\textbf
  {\bibinfo {volume} {427}},\ \bibinfo {pages} {835} (\bibinfo {year}
  {2004})}\BibitemShut {NoStop}%
\bibitem [{\citenamefont {Pimm}\ and\ \citenamefont
  {Lawton}(1980)}]{pimm1980compartments}%
  \BibitemOpen
  \bibfield  {author} {\bibinfo {author} {\bibfnamefont {S.~L.}\ \bibnamefont
  {Pimm}}\ and\ \bibinfo {author} {\bibfnamefont {J.~H.}\ \bibnamefont
  {Lawton}},\ }\href@noop {} {\bibfield  {journal} {\bibinfo  {journal}
  {Journal of Animal Ecology}\ ,\ \bibinfo {pages} {879}} (\bibinfo {year}
  {1980})}\BibitemShut {NoStop}%
\bibitem [{\citenamefont {Woodward}\ \emph {et~al.}(2005)\citenamefont
  {Woodward}, \citenamefont {Ebenman}, \citenamefont {Emmerson}, \citenamefont
  {Montoya}, \citenamefont {Olesen}, \citenamefont {Valido},\ and\
  \citenamefont {Warren}}]{woodward2005body}%
  \BibitemOpen
  \bibfield  {author} {\bibinfo {author} {\bibfnamefont {G.}~\bibnamefont
  {Woodward}}, \bibinfo {author} {\bibfnamefont {B.}~\bibnamefont {Ebenman}},
  \bibinfo {author} {\bibfnamefont {M.}~\bibnamefont {Emmerson}}, \bibinfo
  {author} {\bibfnamefont {J.~M.}\ \bibnamefont {Montoya}}, \bibinfo {author}
  {\bibfnamefont {J.~M.}\ \bibnamefont {Olesen}}, \bibinfo {author}
  {\bibfnamefont {A.}~\bibnamefont {Valido}}, \ and\ \bibinfo {author}
  {\bibfnamefont {P.~H.}\ \bibnamefont {Warren}},\ }\href@noop {} {\bibfield
  {journal} {\bibinfo  {journal} {Trends in Ecology \& Evolution}\ }\textbf
  {\bibinfo {volume} {20}},\ \bibinfo {pages} {402} (\bibinfo {year}
  {2005})}\BibitemShut {NoStop}%
\bibitem [{\citenamefont {Cohen}\ and\ \citenamefont
  {Newman}(1985)}]{cohen-newman-cascade}%
  \BibitemOpen
  \bibfield  {author} {\bibinfo {author} {\bibfnamefont {J.}~\bibnamefont
  {Cohen}}\ and\ \bibinfo {author} {\bibfnamefont {C.}~\bibnamefont {Newman}},\
  }\href@noop {} {\bibfield  {journal} {\bibinfo  {journal} {Proceedings of the
  Royal society of London. Series B. Biological sciences}\ }\textbf {\bibinfo
  {volume} {224}},\ \bibinfo {pages} {421} (\bibinfo {year}
  {1985})}\BibitemShut {NoStop}%
\bibitem [{\citenamefont {Hastie}\ \emph {et~al.}(2009)\citenamefont {Hastie},
  \citenamefont {Tibshirani}, \ and\ \citenamefont {Friedman}}]{hastie2009elements}%
  \BibitemOpen
  \bibfield  {author} {\bibinfo {author} {\bibfnamefont {T.}~\bibnamefont
  {Hastie}}, \bibinfo {author} {\bibfnamefont {R.}~\bibnamefont {Tibshirani}},
  \bibinfo {author} {\bibfnamefont {J.}~\bibnamefont {Friedman}}, 
   }\href@noop {} {\emph
  {\bibinfo {title} {The elements of statistical learning}}},\ Vol.~\bibinfo
  {volume} {2}\ (\bibinfo  {publisher} {Springer},\ \bibinfo {year}
  {2009})\BibitemShut {NoStop}%
\bibitem [{\citenamefont {Earl}\ and\ \citenamefont
  {Deem}(2005)}]{paralleltemp-earl-deem}%
  \BibitemOpen
  \bibfield  {author} {\bibinfo {author} {\bibfnamefont {D.~J.}\ \bibnamefont
  {Earl}}\ and\ \bibinfo {author} {\bibfnamefont {M.~W.}\ \bibnamefont
  {Deem}},\ }\href {\doibase 10.1039/B509983H} {\bibfield  {journal} {\bibinfo
  {journal} {Physical Chemistry Chemical Physics}\ }\textbf {\bibinfo {volume}
  {7}},\ \bibinfo {pages} {3910} (\bibinfo {year} {2005})}\BibitemShut
  {NoStop}%
\bibitem [{\citenamefont {Clauset}\ \emph {et~al.}(2008)\citenamefont
  {Clauset}, \citenamefont {Moore},\ and\ \citenamefont
  {Newman}}]{clauset08-hrg}%
  \BibitemOpen
  \bibfield  {author} {\bibinfo {author} {\bibfnamefont {A.}~\bibnamefont
  {Clauset}}, \bibinfo {author} {\bibfnamefont {C.}~\bibnamefont {Moore}}, \
  and\ \bibinfo {author} {\bibfnamefont {M.~E.~J.}\ \bibnamefont {Newman}},\
  }\href@noop {} {\bibfield  {journal} {\bibinfo  {journal} {Nature}\ }\textbf
  {\bibinfo {volume} {453}},\ \bibinfo {pages} {98} (\bibinfo {year}
  {2008})}\BibitemShut {NoStop}%
\bibitem [{\citenamefont {Stouffer}\ \emph {et~al.}(2005)\citenamefont
  {Stouffer}, \citenamefont {Camacho}, \citenamefont {Guimera}, \citenamefont
  {Ng},\ and\ \citenamefont {Nunes~Amaral}}]{stouffer2005cascade}%
  \BibitemOpen
  \bibfield  {author} {\bibinfo {author} {\bibfnamefont {D.}~\bibnamefont
  {Stouffer}}, \bibinfo {author} {\bibfnamefont {J.}~\bibnamefont {Camacho}},
  \bibinfo {author} {\bibfnamefont {R.}~\bibnamefont {Guimera}}, \bibinfo
  {author} {\bibfnamefont {C.}~\bibnamefont {Ng}}, \ and\ \bibinfo {author}
  {\bibfnamefont {L.}~\bibnamefont {Nunes~Amaral}},\ }\href@noop {} {\bibfield
  {journal} {\bibinfo  {journal} {Ecology}\ }\textbf {\bibinfo {volume} {86}},\
  \bibinfo {pages} {1301} (\bibinfo {year} {2005})}\BibitemShut {NoStop}%
\bibitem [{\citenamefont {Zook}\ \emph {et~al.}(2010)\citenamefont {Zook},
  \citenamefont {Eklof}, \citenamefont {Jacob},\ and\ \citenamefont
  {Allesina}}]{Zook2010}%
  \BibitemOpen
  \bibfield  {author} {\bibinfo {author} {\bibfnamefont {A.~E.}\ \bibnamefont
  {Zook}}, \bibinfo {author} {\bibfnamefont {A.}~\bibnamefont {Eklof}},
  \bibinfo {author} {\bibfnamefont {U.}~\bibnamefont {Jacob}}, \ and\ \bibinfo
  {author} {\bibfnamefont {S.}~\bibnamefont {Allesina}},\ }\href {\doibase
  10.1016/j.jtbi.2010.11.045} {\bibfield  {journal} {\bibinfo  {journal}
  {Journal of Theoretical Biology}\ }\textbf {\bibinfo {volume} {271}},\
  \bibinfo {pages} {106} (\bibinfo {year} {2010})}\BibitemShut {NoStop}%
\bibitem [{\citenamefont {Preston}\ \emph {et~al.}(2014)\citenamefont
  {Preston}, \citenamefont {Jacobs}, \citenamefont {Orlofske},\ and\
  \citenamefont {Johnson}}]{preston2014complex}%
  \BibitemOpen
  \bibfield  {author} {\bibinfo {author} {\bibfnamefont {D.~L.}\ \bibnamefont
  {Preston}}, \bibinfo {author} {\bibfnamefont {A.~Z.}\ \bibnamefont {Jacobs}},
  \bibinfo {author} {\bibfnamefont {S.~A.}\ \bibnamefont {Orlofske}}, \ and\
  \bibinfo {author} {\bibfnamefont {P.~T.}\ \bibnamefont {Johnson}},\
  }\href@noop {} {\bibfield  {journal} {\bibinfo  {journal} {Oecologia}\
  }\textbf {\bibinfo {volume} {174}},\ \bibinfo {pages} {953} (\bibinfo {year}
  {2014})}\BibitemShut {NoStop}%
\bibitem [{\citenamefont {Lafferty}\ \emph {et~al.}(2006)\citenamefont
  {Lafferty}, \citenamefont {Dobson},\ and\ \citenamefont
  {Kuris}}]{parasites-dominate-fw}%
  \BibitemOpen
  \bibfield  {author} {\bibinfo {author} {\bibfnamefont {K.~D.}\ \bibnamefont
  {Lafferty}}, \bibinfo {author} {\bibfnamefont {A.~P.}\ \bibnamefont
  {Dobson}}, \ and\ \bibinfo {author} {\bibfnamefont {A.~M.}\ \bibnamefont
  {Kuris}},\ }\href@noop {} {\bibfield  {journal} {\bibinfo  {journal}
  {Proceedings of the National Academy of Sciences}\ }\textbf {\bibinfo
  {volume} {103}},\ \bibinfo {pages} {11211} (\bibinfo {year}
  {2006})}\BibitemShut {NoStop}%
\bibitem [{\citenamefont {Thieltges}\ \emph {et~al.}(2013)\citenamefont
  {Thieltges}, \citenamefont {Amundsen}, \citenamefont {Hechinger},
  \citenamefont {Johnson}, \citenamefont {Lafferty}, \citenamefont {Mouritsen},
  \citenamefont {Preston}, \citenamefont {Reise}, \citenamefont {Zander},\ and\
  \citenamefont {Poulin}}]{thieltges-paraprey-2013}%
  \BibitemOpen
  \bibfield  {author} {\bibinfo {author} {\bibfnamefont {D.~W.}\ \bibnamefont
  {Thieltges}}, \bibinfo {author} {\bibfnamefont {P.-A.}\ \bibnamefont
  {Amundsen}}, \bibinfo {author} {\bibfnamefont {R.~F.}\ \bibnamefont
  {Hechinger}}, \bibinfo {author} {\bibfnamefont {P.~T.~J.}\ \bibnamefont
  {Johnson}}, \bibinfo {author} {\bibfnamefont {K.~D.}\ \bibnamefont
  {Lafferty}}, \bibinfo {author} {\bibfnamefont {K.~N.}\ \bibnamefont
  {Mouritsen}}, \bibinfo {author} {\bibfnamefont {D.~L.}\ \bibnamefont
  {Preston}}, \bibinfo {author} {\bibfnamefont {K.}~\bibnamefont {Reise}},
  \bibinfo {author} {\bibfnamefont {C.~D.}\ \bibnamefont {Zander}}, \ and\
  \bibinfo {author} {\bibfnamefont {R.}~\bibnamefont {Poulin}},\ }\href
  {\doibase 10.1111/j.1600-0706.2013.00243.x} {\bibfield  {journal} {\bibinfo
  {journal} {Oikos}\ } (\bibinfo {year} {2013}),\
  10.1111/j.1600-0706.2013.00243.x}\BibitemShut {NoStop}%
\bibitem [{\citenamefont {{Newman}}\ and\ \citenamefont
  {{Barkema}}(1999)}]{newman-barkema-MCMC}%
  \BibitemOpen
  \bibfield  {author} {\bibinfo {author} {\bibfnamefont {M.~E.~J.}\
  \bibnamefont {{Newman}}}\ and\ \bibinfo {author} {\bibfnamefont {G.~T.}\
  \bibnamefont {{Barkema}}},\ }\href@noop {} {\emph {\bibinfo {title} {Monte
  Carlo methods in statistical physics.}}}\ (\bibinfo  {publisher} {Oxford: Clarendon
  Press},\ \bibinfo {year} {1999})\BibitemShut {NoStop}%
\end{thebibliography}

\begin{thebibliography}{15}%
\makeatletter
\providecommand \@ifxundefined [1]{%
 \@ifx{#1\undefined}
}%
\providecommand \@ifnum [1]{%
 \ifnum #1\expandafter \@firstoftwo
 \else \expandafter \@secondoftwo
 \fi
}%
\providecommand \@ifx [1]{%
 \ifx #1\expandafter \@firstoftwo
 \else \expandafter \@secondoftwo
 \fi
}%
\providecommand \natexlab [1]{#1}%
\providecommand \enquote  [1]{``#1''}%
\providecommand \bibnamefont  [1]{#1}%
\providecommand \bibfnamefont [1]{#1}%
\providecommand \citenamefont [1]{#1}%
\providecommand \href@noop [0]{\@secondoftwo}%
\providecommand \href [0]{\begingroup \@sanitize@url \@href}%
\providecommand \@href[1]{\@@startlink{#1}\@@href}%
\providecommand \@@href[1]{\endgroup#1\@@endlink}%
\providecommand \@sanitize@url [0]{\catcode `\\12\catcode `\$12\catcode
  `\&12\catcode `\#12\catcode `\^12\catcode `\_12\catcode `\%12\relax}%
\providecommand \@@startlink[1]{}%
\providecommand \@@endlink[0]{}%
\providecommand \url  [0]{\begingroup\@sanitize@url \@url }%
\providecommand \@url [1]{\endgroup\@href {#1}{\urlprefix }}%
\providecommand \urlprefix  [0]{URL }%
\providecommand \Eprint [0]{\href }%
\providecommand \doibase [0]{http://dx.doi.org/}%
\providecommand \selectlanguage [0]{\@gobble}%
\providecommand \bibinfo  [0]{\@secondoftwo}%
\providecommand \bibfield  [0]{\@secondoftwo}%
\providecommand \translation [1]{[#1]}%
\providecommand \BibitemOpen [0]{}%
\providecommand \bibitemStop [0]{}%
\providecommand \bibitemNoStop [0]{.\EOS\space}%
\providecommand \EOS [0]{\spacefactor3000\relax}%
\providecommand \BibitemShut  [1]{\csname bibitem#1\endcsname}%
\let\auto@bib@innerbib\@empty
\bibitem [{\citenamefont {Zander}\ \emph {et~al.}(2007)\citenamefont {Zander},
  \citenamefont {Josten}, \citenamefont {Detloff}, \citenamefont {Poulin},
  \citenamefont {McLaughlin},\ and\ \citenamefont
  {Thieltges}}]{flensburg-fjord-data}%
  \BibitemOpen
  \bibfield  {author} {\bibinfo {author} {\bibfnamefont {C.~D.}\ \bibnamefont
  {Zander}}, \bibinfo {author} {\bibfnamefont {N.}~\bibnamefont {Josten}},
  \bibinfo {author} {\bibfnamefont {K.~C.}\ \bibnamefont {Detloff}}, \bibinfo
  {author} {\bibfnamefont {R.}~\bibnamefont {Poulin}}, \bibinfo {author}
  {\bibfnamefont {J.~P.}\ \bibnamefont {McLaughlin}}, \ and\ \bibinfo {author}
  {\bibfnamefont {D.~W.}\ \bibnamefont {Thieltges}},\ }\href@noop {} {\bibfield
   {journal} {\bibinfo  {journal} {Ecology}\ }\textbf {\bibinfo {volume} {92}}
  (\bibinfo {year} {2007})}\BibitemShut {NoStop}%
\bibitem [{\citenamefont {Williams}\ \emph {et~al.}(2010)\citenamefont
  {Williams}, \citenamefont {Anandanadesan},\ and\ \citenamefont
  {Purves}}]{PNM2010}%
  \BibitemOpen
  \bibfield  {author} {\bibinfo {author} {\bibfnamefont {R.~J.}\ \bibnamefont
  {Williams}}, \bibinfo {author} {\bibfnamefont {A.}~\bibnamefont
  {Anandanadesan}}, \ and\ \bibinfo {author} {\bibfnamefont {D.}~\bibnamefont
  {Purves}},\ }\href {\doibase 10.1371/journal.pone.0012092} {\bibfield
  {journal} {\bibinfo  {journal} {PLOS ONE}\ }\textbf {\bibinfo {volume} {5}},\
  \bibinfo {pages} {e12092} (\bibinfo {year} {2010})}\BibitemShut {NoStop}%
\bibitem [{\citenamefont {Williams}\ and\ \citenamefont
  {Purves}(2011)}]{PNM2011}%
  \BibitemOpen
  \bibfield  {author} {\bibinfo {author} {\bibfnamefont {R.~J.}\ \bibnamefont
  {Williams}}\ and\ \bibinfo {author} {\bibfnamefont {D.~W.}\ \bibnamefont
  {Purves}},\ }\href@noop {} {\bibfield  {journal} {\bibinfo  {journal}
  {Ecology}\ }\textbf {\bibinfo {volume} {92}},\ \bibinfo {pages} {1849}
  (\bibinfo {year} {2011})}\BibitemShut {NoStop}%
\bibitem [{\citenamefont {Williams}\ and\ \citenamefont
  {Martinez}(2000)}]{williams-martinez-niche-2000}%
  \BibitemOpen
  \bibfield  {author} {\bibinfo {author} {\bibfnamefont {R.~J.}\ \bibnamefont
  {Williams}}\ and\ \bibinfo {author} {\bibfnamefont {N.~D.}\ \bibnamefont
  {Martinez}},\ }\href@noop {} {\bibfield  {journal} {\bibinfo  {journal}
  {Nature}\ }\textbf {\bibinfo {volume} {404}},\ \bibinfo {pages} {180}
  (\bibinfo {year} {2000})}\BibitemShut {NoStop}%
\bibitem [{\citenamefont {Dunne}\ \emph {et~al.}(2013)\citenamefont {Dunne},
  \citenamefont {Lafferty}, \citenamefont {Dobson}, \citenamefont {Hechinger},
  \citenamefont {Kuris}, \citenamefont {Martinez}, \citenamefont {McLaughlin},
  \citenamefont {Mouritsen}, \citenamefont {Poulin}, \citenamefont {Reise}
  \emph {et~al.}}]{dunne-2013-parafw}%
  \BibitemOpen
  \bibfield  {author} {\bibinfo {author} {\bibfnamefont {J.~A.}\ \bibnamefont
  {Dunne}}, \bibinfo {author} {\bibfnamefont {K.~D.}\ \bibnamefont {Lafferty}},
  \bibinfo {author} {\bibfnamefont {A.~P.}\ \bibnamefont {Dobson}}, \bibinfo
  {author} {\bibfnamefont {R.~F.}\ \bibnamefont {Hechinger}}, \bibinfo {author}
  {\bibfnamefont {A.~M.}\ \bibnamefont {Kuris}}, \bibinfo {author}
  {\bibfnamefont {N.~D.}\ \bibnamefont {Martinez}}, \bibinfo {author}
  {\bibfnamefont {J.~P.}\ \bibnamefont {McLaughlin}}, \bibinfo {author}
  {\bibfnamefont {K.~N.}\ \bibnamefont {Mouritsen}}, \bibinfo {author}
  {\bibfnamefont {R.}~\bibnamefont {Poulin}}, \bibinfo {author} {\bibfnamefont
  {K.}~\bibnamefont {Reise}},  \emph {et~al.},\ }\href@noop {} {\bibfield
  {journal} {\bibinfo  {journal} {PLOS biology}\ }\textbf {\bibinfo {volume}
  {11}},\ \bibinfo {pages} {e1001579} (\bibinfo {year} {2013})}\BibitemShut
  {NoStop}%
\bibitem [{\citenamefont {Jiang}(2010)}]{largesampletechniques}%
  \BibitemOpen
  \bibfield  {author} {\bibinfo {author} {\bibfnamefont {J.}~\bibnamefont
  {Jiang}},\ }in\ \href {http://dx.doi.org/10.1007/978-1-4419-6827-2\_11}
  {\emph {\bibinfo {booktitle} {Large Sample Techniques for Statistics}}},\
  \bibinfo {series} {Springer Texts in Statistics}, Vol.~\bibinfo {volume}
  {0},\ \bibinfo {editor} {edited by\ \bibinfo {editor} {\bibfnamefont
  {G.}~\bibnamefont {Casella}}, \bibinfo {editor} {\bibfnamefont
  {S.}~\bibnamefont {Fienberg}}, \ and\ \bibinfo {editor} {\bibfnamefont
  {I.}~\bibnamefont {Olkin}}}\ (\bibinfo  {publisher} {Springer New York},\
  \bibinfo {year} {2010})\ pp.\ \bibinfo {pages} {357--391},\ \bibinfo {note}
  {10.1007/978-1-4419-6827-2\_11}\BibitemShut {NoStop}%
\bibitem [{\citenamefont {Clauset}\ \emph {et~al.}(2008)\citenamefont
  {Clauset}, \citenamefont {Moore},\ and\ \citenamefont
  {Newman}}]{clauset08-hrg}%
  \BibitemOpen
  \bibfield  {author} {\bibinfo {author} {\bibfnamefont {A.}~\bibnamefont
  {Clauset}}, \bibinfo {author} {\bibfnamefont {C.}~\bibnamefont {Moore}}, \
  and\ \bibinfo {author} {\bibfnamefont {M.~E.~J.}\ \bibnamefont {Newman}},\
  }\href@noop {} {\bibfield  {journal} {\bibinfo  {journal} {Nature}\ }\textbf
  {\bibinfo {volume} {453}},\ \bibinfo {pages} {98} (\bibinfo {year}
  {2008})}\BibitemShut {NoStop}%
\bibitem [{\citenamefont {Lobo}\ \emph {et~al.}(2008)\citenamefont {Lobo},
  \citenamefont {Jim{\'e}nez-Valverde},\ and\ \citenamefont
  {Real}}]{auc-misleading}%
  \BibitemOpen
  \bibfield  {author} {\bibinfo {author} {\bibfnamefont {J.~M.}\ \bibnamefont
  {Lobo}}, \bibinfo {author} {\bibfnamefont {A.}~\bibnamefont
  {Jim{\'e}nez-Valverde}}, \ and\ \bibinfo {author} {\bibfnamefont
  {R.}~\bibnamefont {Real}},\ }\href {\doibase
  10.1111/j.1466-8238.2007.00358.x} {\bibfield  {journal} {\bibinfo  {journal}
  {Global Ecology and Biogeography}\ }\textbf {\bibinfo {volume} {17}},\
  \bibinfo {pages} {145} (\bibinfo {year} {2008})}\BibitemShut {NoStop}%
\bibitem [{\citenamefont {Earl}\ and\ \citenamefont
  {Deem}(2005)}]{paralleltemp-earl-deem}%
  \BibitemOpen
  \bibfield  {author} {\bibinfo {author} {\bibfnamefont {D.~J.}\ \bibnamefont
  {Earl}}\ and\ \bibinfo {author} {\bibfnamefont {M.~W.}\ \bibnamefont
  {Deem}},\ }\href {\doibase 10.1039/B509983H} {\bibfield  {journal} {\bibinfo
  {journal} {Physical Chemistry Chemical Physics}\ }\textbf {\bibinfo {volume}
  {7}},\ \bibinfo {pages} {3910} (\bibinfo {year} {2005})}\BibitemShut
  {NoStop}%
\bibitem [{\citenamefont {{Newman}}\ and\ \citenamefont
  {{Barkema}}(1999)}]{newman-barkema-MCMC}%
  \BibitemOpen
  \bibfield  {author} {\bibinfo {author} {\bibfnamefont {M.~E.~J.}\
  \bibnamefont {{Newman}}}\ and\ \bibinfo {author} {\bibfnamefont {G.~T.}\
  \bibnamefont {{Barkema}}},\ }\href@noop {} {\emph {\bibinfo {title} {Monte
  Carlo methods in statistical physics.~Oxford
  : Clarendon Press, 1999.}}}\ (\bibinfo  {publisher} {Oxford: Clarendon
  Press},\ \bibinfo {year} {1999})\BibitemShut {NoStop}%
\bibitem [{Note1()}]{Note1}%
  \BibitemOpen
  \bibinfo {note} {In this section and section S4.2, all results shown are
  calculated from optima found from each of 100 different parallel tempering
  runs, as used in the main text.}\BibitemShut {Stop}%
\bibitem [{\citenamefont {Zook}\ \emph {et~al.}(2010)\citenamefont {Zook},
  \citenamefont {Eklof}, \citenamefont {Jacob},\ and\ \citenamefont
  {Allesina}}]{Zook2010}%
  \BibitemOpen
  \bibfield  {author} {\bibinfo {author} {\bibfnamefont {A.~E.}\ \bibnamefont
  {Zook}}, \bibinfo {author} {\bibfnamefont {A.}~\bibnamefont {Eklof}},
  \bibinfo {author} {\bibfnamefont {U.}~\bibnamefont {Jacob}}, \ and\ \bibinfo
  {author} {\bibfnamefont {S.}~\bibnamefont {Allesina}},\ }\href {\doibase
  10.1016/j.jtbi.2010.11.045} {\bibfield  {journal} {\bibinfo  {journal}
  {Journal of Theoretical Biology}\ }\textbf {\bibinfo {volume} {271}},\
  \bibinfo {pages} {106} (\bibinfo {year} {2010})}\BibitemShut {NoStop}%
\bibitem [{\citenamefont {Cattin}\ \emph {et~al.}(2004)\citenamefont {Cattin},
  \citenamefont {Bersier}, \citenamefont {Bana{\v{s}}ek-Richter}, \citenamefont
  {Baltensperger},\ and\ \citenamefont {Gabriel}}]{cattin2004phylogenetic}%
  \BibitemOpen
  \bibfield  {author} {\bibinfo {author} {\bibfnamefont {M.-F.}\ \bibnamefont
  {Cattin}}, \bibinfo {author} {\bibfnamefont {L.-F.}\ \bibnamefont {Bersier}},
  \bibinfo {author} {\bibfnamefont {C.}~\bibnamefont {Bana{\v{s}}ek-Richter}},
  \bibinfo {author} {\bibfnamefont {R.}~\bibnamefont {Baltensperger}}, \ and\
  \bibinfo {author} {\bibfnamefont {J.-P.}\ \bibnamefont {Gabriel}},\
  }\href@noop {} {\bibfield  {journal} {\bibinfo  {journal} {Nature}\ }\textbf
  {\bibinfo {volume} {427}},\ \bibinfo {pages} {835} (\bibinfo {year}
  {2004})}\BibitemShut {NoStop}%
\bibitem [{\citenamefont {Warren}\ \emph {et~al.}(2010)\citenamefont {Warren},
  \citenamefont {Pascual}, \citenamefont {Lafferty},\ and\ \citenamefont
  {Kuris}}]{inverse-niche-model}%
  \BibitemOpen
  \bibfield  {author} {\bibinfo {author} {\bibfnamefont {C.~P.}\ \bibnamefont
  {Warren}}, \bibinfo {author} {\bibfnamefont {M.}~\bibnamefont {Pascual}},
  \bibinfo {author} {\bibfnamefont {K.~D.}\ \bibnamefont {Lafferty}}, \ and\
  \bibinfo {author} {\bibfnamefont {A.~M.}\ \bibnamefont {Kuris}},\ }\href@noop
  {} {\bibfield  {journal} {\bibinfo  {journal} {Theoretical Ecology}\ }\textbf
  {\bibinfo {volume} {3}},\ \bibinfo {pages} {285} (\bibinfo {year}
  {2010})}\BibitemShut {NoStop}%
\bibitem [{\citenamefont {Cirtwill}\ and\ \citenamefont
  {Stouffer}(2015)}]{cirtwill2015concomitant}%
  \BibitemOpen
  \bibfield  {author} {\bibinfo {author} {\bibfnamefont {A.~R.}\ \bibnamefont
  {Cirtwill}}\ and\ \bibinfo {author} {\bibfnamefont {D.~B.}\ \bibnamefont
  {Stouffer}},\ }\href {\doibase 10.1111/1365-2656.12323} {\bibfield  {journal}
  {\bibinfo  {journal} {Journal of Animal Ecology}\ }\textbf {\bibinfo {volume}
  {84}},\ \bibinfo {pages} {734} (\bibinfo {year} {2015})}\BibitemShut
  {NoStop}%
\end{thebibliography}
\end{document}